\documentclass[12pt,a4paper]{article}

\usepackage{amsthm,amssymb,latexsym,amscd,amsmath,enumerate,amsfonts,mathrsfs}
\usepackage[latin1]{inputenc}
\usepackage[english]{babel}
\usepackage[T1]{fontenc}
\usepackage{graphics,graphicx}
\usepackage{cite}
\usepackage[usenames,svgnames,dvipsnames]{xcolor}
\usepackage[pdftex,plainpages=false,pdfpagelabels,pagebackref,colorlinks=true,citecolor=ForestGreen,linkcolor=NavyBlue,urlcolor=DarkRed,filecolor=green,bookmarksopen=true]{hyperref}
\usepackage[font=small,format=plain,labelfont=bf,up]{caption}
\usepackage[a4paper,top=2.54cm,bottom=2.4cm,left=2.3cm,right=2.3cm]{geometry}

\usepackage{fancyhdr}
\begin{document}

%
%


%
%


\def \beq {\begin{equation}}
\def \eeq {\end{equation}}
\def \bearr {\begin{eqnarray}}
\def \eearr {\end{eqnarray}}
\def \l {'}
\def \q {?}
\def \up {|}
\def \bN {{\bf \nabla}}
\def \bT {{\bf T}}
\def \bL {{\bf L}}
\def \dL {\textit{L}}
\def \bv {{\bf v}}
\def \bu {{\bf u}}
\def \ba {{\bf a}}
\def \br {{\bf r}}
\def \be {{\bf e}}
\def \bw {{\bf w}}
\def \bd {{\bf d}}
\def \bp {{\bf p}}
\def \bx {\mathbf{x}}
\def \bk {\mathbf{k}}
\def \bP {\mathbf{P}}
\def \bl {\mathbf{l}}
\def \bm {\mathbf{m}}
\def \del {\partial}
\def \calL {\mathcal{L}}
\def \calH {\mathcal{H}}
\def \calD {\mathcal{D}}
\def \calP {\mathcal{P}}
\def \calM {\mathcal{M}}
\def \calA {\mathcal{A}}
\def \calR {\mathcal{R}} 
\def \calS {\mathcal{S}}
\def \calF {\mathcal{F}}
\def \calO {\mathcal{O}}
\def \tinymu {{\scripscriptstyle \mu}}
\def \tinynu {{\scripscriptstyle \nu}}
\def \qhat {\hat{q}}
\def \phat {\hat{p}}
\def \ahat {\hat{a}}
\def \adagger {\hat{a}^{\dagger}}
\def \aprhat {\hat{a}'}
\def \aprdagger {\hat{a}'^{\dagger}}
\def \phihat {\hat{\phi}}
\def \pihat {\hat{\pi}}
\def \nhat {\hat{n}}
\def \vac {| 0 \rangle}
\def \vacpr {| 0' \rangle}
\def \d {\mathrm{d}}
\def \epsbar {\bar{\epsilon}}
\def \redshift {\emph{redshift}}
\def \Schw {{Schwarzschild}}
\def \rank {\mathrm{rank} \,}
\def \RN {Reissner-Nordstr�m}
\def \zetabar {\bar{\zeta}}
\def \Lie {\mathscr{L}}
\def \epsb {\varepsilon}
\def \scri {\mathscr{I}}
\def \Lbar {\bar{L}}
\def \Wbar {\bar{W}}
\def \delbar {\bar{\partial}}
\def \mbar {\bar{m}}
\def \Dscr {\mathscr{D}}
\def \deltabar {\bar{\delta}}

\newtheorem{thm}{Theorem}[section]
\newtheorem{lemma}[thm]{Lemma}
\newtheorem{cor}[thm]{Corollary}
\newtheorem{prop}[thm]{Proposition}
\theoremstyle{definition}
\newtheorem{defn}[thm]{Definition}
\newtheorem{eg}[thm]{Example}

%
%

\setcounter{tocdepth}{2}
\fancyhead{}
\fancyhead[LO,LE]{{\footnotesize\rightmark}}
\fancyhead[RO,RE]{\thepage}


\title{\bf Algebraically special solutions in AdS/CFT}
\author{Gabriel Bernardi de Freitas\thanks{\texttt{gbf23@cam.ac.uk}} {}  and Harvey S. Reall\thanks{\texttt{hsr1000@cam.ac.uk}} \\
        \hfill \\
        \normalsize{Department of Applied Mathematics and Theoretical Physics, University of Cambridge} \\
        \normalsize{Centre for Mathematical Sciences, Wilberforce Road, Cambridge CB3 0WA, United Kingdom}}
\date{\today}
\maketitle

\begin{abstract}
We investigate the AdS/CFT interpretation of the class of algebraically special solutions of Einstein gravity with a negative cosmological constant. Such solutions describe a CFT living in a $2+1$ dimensional time-dependent geometry that, generically, has no isometries. The algebraically special condition implies that the expectation value of the CFT energy-momentum tensor is a local function of the boundary metric. When such a spacetime is slowly varying, the fluid/gravity approximation is valid and one can read off the values of certain higher order transport coefficients. To do this, we introduce a formalism for studying conformal, relativistic fluids in $2+1$ dimensions that reduces everything to the manipulation of scalar quantities.

\end{abstract}


%
%

\section{Introduction}

The AdS/CFT correspondence relates a gravitational theory in $d+1$ dimensions to a CFT in $d$ dimensions. Various explicit solutions of the Einstein equations with a negative cosmological constant have been studied in this context, for example stationary black hole solutions are known to describe thermal equilibrium states of the CFT. An interesting class of solutions of the $3+1$ dimensional Einstein equations are the solutions with an algebraically special Weyl tensor. This includes black hole solutions such as Kerr-AdS but also much more general solutions which have free functional degrees of freedom and no isometries. In this paper we will study the AdS/CFT interpretation of such solutions. 

The Goldberg-Sachs theorem implies that the algebraically special property is equivalent to the existence of a null geodesic congruence with vanishing shear. In this paper we will focus on spacetimes for which this congruence also has vanishing rotation but non-vanishing expansion. This defines the 
{\it Robinson-Trautman} (RT) family of algebraically special solutions \cite{RobinsonTrautman1962}. A general member of the RT family is a time-dependent deformation of the Schwarzschild-AdS solution. As we will explain in Section \ref{RT-section}, these solutions have a simple explicit form depending on one function $\lambda$ of three coordinates. This function is constrained to satisfy a certain parabolic PDE: the RT equation. 

The conformal boundary of a RT solution is a 2+1 dimensional spacetime with metric
\beq
 ds^2 = - \d t^2 + e^{2\lambda(t,x)} \bar{g}_{ij}(x) \d x^i \d x^j
\eeq
where $\bar{g}_{ij}$ is a two-dimensional metric of constant curvature and $\lambda$ is the function mentioned above. In general this metric has no symmetries. One can regard it as an inhomogeneous cosmological spacetime. The CFT lives in this spacetime. Using AdS/CFT we determine the expectation value of the CFT energy-momentum tensor in a state dual to a RT solution. We find that this can be written very simply as the sum of a perfect fluid part, with the fluid at rest in the above coordinates, and a 3-derivative term constructed from the curvature of the boundary metric.

From a CFT perspective, this result is surprising because it is a {\it local} function of the boundary geometry, i.e., $\langle T_{ab}(t,x) \rangle$ depends only on the spacetime geometry at the point $(t,x)$ and not, as would generically be the case, on the geometry at other points e.g. those in the past light-cone of this point. This result can be attributed to the fact that the bulk spacetime contains only {\it ingoing} radiation: the algebraically special property implies that there is no scattering in the bulk and hence information does not propagate from the boundary at an early time back to the boundary at a later time.

When the CFT is in a state in which observables vary slowly compared to the microscopic scale of the theory (e.g. set by the mean free path) then it can be described by a hydrodynamic derivative expansion. In the simplest situation of an uncharged fluid, the fluid is described by slowly varying temperature and velocity fields. The fluid stress tensor is expressed as an expansion in derivatives of these quantities. The coefficients in this expansion are known as transport coefficients. The {\it fluid/gravity correspondence} of Ref. \cite{BhattacharyyaEtAl2008a} (for a recent review see \cite{HubenyMinwallaRangamani2011}) postulates the form of the bulk metric dual to a general motion of the fluid. This metric takes the form of an infinite derivative expansion, whose form has been determined explicitly up to two derivatives. The bulk Einstein equation and the AdS/CFT correspondence then determine the transport coefficients of the dual fluid.

The metric of the fluid/gravity correspondence is sufficiently complicated that it seems worth looking at particular cases in which it simplifies. One practical reason for doing this is that it might be possible to determine the derivative expansion explicitly to higher order in a particular case than has been achieved in the general case. If the particular case is "sufficiently general", then it might be possible to determine some of the higher order transport coefficients this way. (This possibility was also discussed in Ref. \cite{Mukhopadhyay:2013gja}, which considered a class of stationary bulk solutions for which $\langle T_{ab} \rangle$ takes a perfect fluid form.)

In this paper, we will study the RT solution using the methods of the fluid/gravity correspondence. In this case we know the expectation value of the CFT energy momentum tensor exactly so we can expand it in derivatives to an arbitrarily high order. Since the solution contains free functional degrees of freedom, it seems likely that it is general enough for this procedure to determine certain higher order transport coefficients. 

To define transport coefficients, one must write out the derivative expansion of the energy-momentum tensor including all possible terms that can arise at each order in derivatives, modulo lower-order equations of motion and geometrical identities. Identifying all possible terms becomes complicated beyond second order. To simplify this problem, we introduce a new formalism for studying conformal relativistic fluid mechanics in 2+1 dimensions. This is inspired by the Geroch-Held-Penrose formalism in general relativity \cite{GerochHeldPenrose1973}. In our formalism, everything is reduced to the manipulation of scalar quantities. This makes the classification of higher derivative terms much more straightforward than in a tensorial approach.

Applying this formalism to the energy-momentum tensor obtained from the RT solutions, we find that certain transport coefficients associated to four- and six-derivative terms are determined uniquely. However, somewhat disappointingly, it turns out that transport coefficients associated to three-derivative terms are not constrained. We also study the entropy current defined by these solutions, and find that some higher-order curvature terms have no contribution to the divergence of the entropy current. In addition, our formalism enables us to determine uniquely the coefficients of some higher-order terms in the entropy current.

Another nice example is the Kerr-AdS bulk metric. In this case, the CFT lives in the Einstein static universe $\mathbb{R} \times S^2$. 
For a large black hole, the hydrodynamic description of the CFT should be valid. In Refs. \cite{AwadJohnson2000,BhattacharyyaEtAl2008b}, the CFT energy-momentum tensor was determined. It was found to take the form of a perfect fluid rotating rigidly around the boundary sphere. The fluid has vanishing shear but non-vanishing rotation. This seems surprising: the perfect fluid form should be the leading order result but one would have expected higher derivative corrections to the energy-momentum tensor. In particular one might have expected terms constructed from the fluid rotation. Using our formalism, we again find that this result does not constrain any three-derivative terms but it does constrain transport coefficients at four derivatives. 

The RT solutions have a null geodesic congruence with vanishing shear and rotation. There exists a larger family of algebraically special solutions for which the shear vanishes but not the rotation. The dependence of the bulk metric on a "radial" coordinate (an affine parameter along the geodesics) is known explicitly, with the Einstein equation reducing to certain PDEs constraining the dependence on the other coordinates (i.e. the "boundary" coordinates). We briefly comment below on some results on the dual CFT interpretation of this family of solutions. In this case, the CFT lives in a 2+1 dimensional spacetime which is rotating: 
\beq
 ds^2 = -(\d t + a_i(t,x)\,  \d x^i)^2 +  e^{2\lambda(t,x)} \bar{g}_{ij}(x) \d x^i \d x^j.
\eeq
Once again we find that the CFT energy-momentum tensor can be written as the sum of a perfect fluid part, with the fluid at rest in the above coordinates, and the same three-derivative term constructed from the curvature of the boundary metric as we discussed above. 

This paper is organised as follows. We review some properties of RT spacetimes in Section \ref{RT-section} and study the energy-momentum tensor of their dual CFT state. 
 As we are interested in studying also the dual CFT stress tensor in view of the fluid/gravity correspondence, in Section \ref{section-conf-fluids-3-dim} we introduce our aforemontioned formalism to study conformal fluids in 2+1 dimensions. We apply the formalism to the RT case in Section \ref{section-RT-fluid-gravity} and to the Kerr-AdS case in Section \ref{Kerr-AdS-section}. In Section \ref{conclusions} we discuss algebraically special solutions with non-vanishing rotation and suggest possible future directions for research.

%
%

\section{Robinson-Trautman solutions}
\label{RT-section}

\subsection{Properties of RT solutions}

Robinson-Trautman spacetimes \cite{RobinsonTrautman1962} are an important class of exact solutions of the Einstein equations. They are defined by the existence of a geodesic, shear-free, twist-free but expanding null congruence. According to the Goldberg-Sachs theorem, all such spacetimes are algebraically special in the vacuum case, with the defining null congruence being a repeated principal null direction. 

In standard form, the metric satisfying the above properties with a negative cosmological constant $\Lambda = -3/l^2$ can be written as \cite{ExactSolnsBook2003,GriffithsPodolsky2012}
\beq
ds^2 = - \Phi \d u^2 - 2 \d u \d r + r^2 g_{(2)} \qquad g_{(2)} = \frac{2}{P(u,\zeta,\zetabar)^2} \d \zeta \d \zetabar,
\label{RT-metric}
\eeq
with
\beq
\Phi = K - 2r \del_u \ln P - \frac{2m}{r} + \frac{r^2}{l^2}.
\eeq
where $m$ is a constant,
\beq
 K = \Delta \ln P \qquad \Delta = 2P^2 \del_{\zeta} \del_{\zetabar}
\eeq
$\Delta$ is the Laplacian of a two-dimensional manifold with metric $g_{(2)}$ and $K$ is the Gaussian curvature of this metric (i.e. half the Ricci scalar). The function $P(u,\zeta,\zetabar)$, must obey the \emph{Robinson-Trautman equation},
\beq
\partial_u \ln P = - \frac{1}{12m} \Delta K
\label{RT-eqn}
\eeq
which guarantees that the metric (\ref{RT-metric}) is a solution of the vacuum Einstein equations
\beq
G_{ab} = \frac{3}{l^2} g_{ab}.
\eeq
Note that the RT equation is independent of the cosmological constant. For simplicity, we will from now on choose units such that the AdS scale is set to unity, $l = 1$. 

We will assume that the coordinates $\zeta,\zetabar$ parameterize a compact two-dimensional manifold ${}^{(2)}M$, e.g $S^2$ or $T^2$. The coordinates $(u,r,\zeta,\zetabar)$ are analogous to outgoing Eddington-Finkelstein coordinates for the Schwarzschild solution with $\partial/\partial r$ tangent to the (affinely parameterized) outgoing null geodesics with vanishing shear and rotation, and positive expansion. There is a curvature singularity at $r=0$.

The above metric admits a timelike conformal boundary at $r=\infty$ with topology $\mathbb{R} \times {}^{(2)}M$. The boundary metric can be chosen to be
\beq
 ds^2 = -\d u^2 + g_{(2)}
\eeq
which can be interpreted as an inhomogeneous cosmological spacetime. In general, the boundary metric is not conformally flat so the solution is not asymptotically AdS in the usual sense\footnote{In contrast, RT solutions with vanishing cosmological constant are asymptotically flat at future null infinity.}. According to AdS/CFT, the dual CFT lives in this 2+1 dimensional geometry. 

The Schwarzschild-AdS solution is recovered from the RT metric by taking $g_{(2)}$ to be a time-independent metric of constant unit curvature:
\beq
 P = P_0(\zeta,\zetabar) \equiv 1 + \frac{K}{2} \zeta \zetabar.
\eeq
with $K \in \{1,0,-1\}$ corresponding to spherical, planar, or hyperbolic symmetry.

In the general case it is convenient to write
\beq
P = e^{-\lambda(u,\zeta,\zetabar)} P_0(\zeta,\zetabar)
\label{P-f-def}
\eeq
so the RT equation becomes an equation for $\lambda$. The parabolic nature of the RT equation implies that it comes with a preferred direction of time \cite{Chrusciel1991}. We assume that $m>0$. Then, given initial data specified by a smooth function $\lambda(u_0,\zeta,\zetabar)$, there exists a unique solution of the RT equation for $u \ge u_0$. Moreover, this solution is {\it analytic} in $\zeta,\zetabar$ for all $u>u_0$.\footnote{More precisely: if we write $\zeta = x+iy$ then the solution is a real analytic function of $x,y$.} Hence if we are given smooth but non-analytic data $\lambda(u_0,\zeta,\zetabar)$ then there will exist no corresponding solution of the RT solution for $u<u_0$. 

Given arbitrary smooth initial data $\lambda(u_0,\zeta,\zetabar)$, it has been shown that the corresponding solution $\lambda(u,\zeta,\zetabar)$ converges exponentially fast to a constant $\lambda_\infty$ as $u \rightarrow \infty$  \cite{Chrusciel1991}. This holds for ${}^{(2)}M$ of arbitrary genus. The RT equation is volume preserving, i.e., the volume of ${}^{(2)}M$ with metric $g_{(2)}$ is constant. This determines the value of the constant $\lambda_\infty$. By a rescaling of the coordinates $u,r$ one can arrange that $\lambda_\infty=0$. The rate of convergence is determined by the first non-zero eigenvalue $\nu_1$ of the Laplacian of the metric corresponding to $P_0$ on ${}^{(2)}M$:
\beq
\lambda = {\cal O}(e^{-\nu_1 u/(12m)}).
\eeq
Since $\lambda=0$ corresponds to the Schwarzschild-AdS solution, we can say that RT solutions "settle down" to the Schwarzschild-AdS solution with mass parameter $m$ as $u \rightarrow \infty$. Since $u=\infty$ corresponds to the future event horizon ${\cal H}^+$ of the Schwarzschild-AdS solution, it is natural to try to extend the RT spacetime across the null hypersurface $u=\infty$ by gluing to it the  part of the Schwarzschild-AdS solution that lies beyond ${\cal H}^+$. 

\begin{figure}[ht]

\begin{center}

\includegraphics[width=9cm]{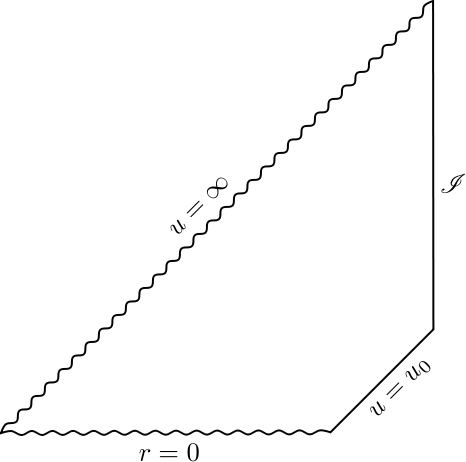}

%
%
%
%

\end{center}

\caption{Conformal structure of a RT spacetime with negative cosmological constant. The solution exists to the future of the null hypersurface $u = u_0$, has a timelike infinity, and approaches the {\Schw}-AdS solution as $u \rightarrow \infty$. However, for large $m$, the hypersurface $u = \infty$ is actually a null singularity. There is also a curvature singularity at $r = 0$.}
\label{fig:RT}

\end{figure}

In the case of vanishing cosmological constant, it has been shown \cite{ChruscielSingleton1992} (cf. also \cite{BicakPodolsky1997}) that the resulting spacetime is not smooth at $u=\infty$: the metric is $C^5$ but not $C^6$ there\footnote{However, there are other extensions that are $C^{117}$ and, generically, this is the smoothest possible \cite{ChruscielSingleton1992, BicakPodolsky1997}.}. This level of smoothness seems physically acceptable. A negative cosmological constant, however, reduces the smoothness of the extension. In particular, for $m^2 > 4/27$, there exists no $C^1$ extension \cite{BicakPodolsky1997}. We will be interested mainly in the case of large $m$ (which is required for validity of the fluid/gravity correspondence) so we will assume that no $C^1$ extension exists, i.e., $u=\infty$ corresponds to a null singularity. This gives the Penrose diagram of Fig. \ref{fig:RT}.

\subsection{Time-reversed RT solution}

The above solution is physically unsatisfactory because of the singularity at $u=\infty$. We can circumvent this problem by applying time reversal: set $u=-t$ to bring the metric to the form
\beq
 ds^2 = - \Phi \d t^2 + 2 \d t \d r + r^2 g_{(2)} \qquad g_{(2)} = \frac{2}{P(t,\zeta,\zetabar)^2} \d \zeta \d \zetabar,
\label{RT-metric2}
\eeq
\beq
\Phi =K + 2r \del_t \ln P - \frac{2m}{r} + r^2.
\eeq
and the RT equation is
\beq
 \partial_t \ln P =  \frac{1}{12m} \Delta K
\label{RTtimereversed}
\eeq
with $K = \Delta \ln P$ as before. Choosing the time orientation so that $-\partial/\partial r$ is future directed, curves of constant $t, \zeta,\zetabar$ are now {\it ingoing} null geodesics with vanishing rotation and shear. 

The above metric admits a timelike conformal boundary as $r \rightarrow \infty$. One can choose the conformal frame so that the boundary metric is (writing $P=e^{-\lambda} P_0$ as above)
\beq
\label{boundary}
 ds_3^2 = -  dt^2 + \frac{2}{P(t,\zeta,\zetabar)^2} d\zeta d\zetabar = -dt^2 + e^{2\lambda(t,\zeta,\zetabar)} \hat{g}_{(2)},
\eeq
where $\hat{g}_{(2)}$ is a metric of constant curvature with $K \in \{1,0,-1\}$. 

\begin{figure}[ht]

\begin{center}

\includegraphics[width=9cm]{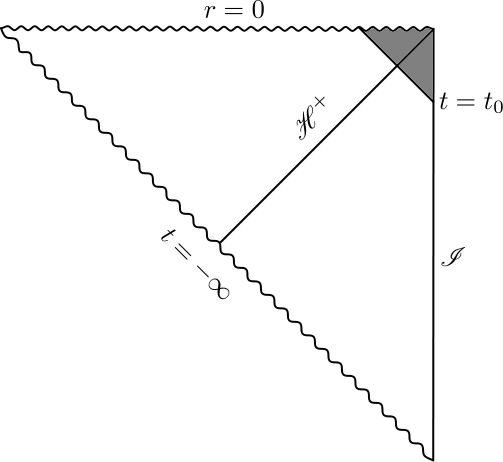}


%
%

\end{center}

\caption{Penrose diagram for the time-reversed (extended) RT solution. The shaded region represents the extension to the future of the hypersurface $t = t_0$, where the spacetime is no longer RT. The location of the event horizon $\mathscr{H}^+$ is also shown. The solution approaches Schwarzschild-AdS as $t\rightarrow -\infty$ but (for large $m$) there exists no $C^1$ extension across this null surface.}

\label{RT:fig:time-rev}

\end{figure}

The time-reversed RT equation (\ref{RTtimereversed}) can only be solved {\it backwards} in time: if $\lambda$ is specified at time $t=t_0$ then there exists a unique solution for $t \le t_0$. This solution has $\lambda \rightarrow {\rm constant}$ as $t \rightarrow -\infty$. Hence, in the bulk, the solution exists to the past of the null hypersurface $t=t_0$ and approaches the Schwarzschild-AdS solution as $t \rightarrow -\infty$. However, since there exists no $C^1$ extension across the null hypersurface $t=-\infty$ for large $m$, this surface is really a null singularity, see Fig. \ref{RT:fig:time-rev}.


The null hypersurface $t=t_0$ is a future boundary of the bulk. It seems very likely that the bulk solution can be extended to the future of this null hypersurface. An extension could be constructed by specifying initial data in the bulk on the $t=t_0$ hypersurface to be that given by the RT solution. If this is supplemented with a specification of the conformal boundary metric for $t \ge t_0$ then there should exist a unique bulk solution to the future of $t=t_0$, although this will not be a RT metric. If the boundary metric is chosen so that it is smooth at $t=t_0$ then the bulk solution should be smooth at $t=t_0$. For example, one could define the conformal boundary metric to take the form (\ref{boundary}) for $t>t_0$ with the function $\lambda$ chosen to match smoothly onto the RT solution at $t = t_0$. Taking $\lambda$ to approach a constant sufficiently rapidly as $t \rightarrow \infty$ one would expect the bulk solution to settle down to Schwarzschild-AdS at late time. This spacetime will then possess an event horizon as shown in Fig. \ref{RT:fig:time-rev}. We will refer to this spacetime as an \emph{extended RT solution}. 

\subsection{CFT interpretation}

We can now discuss the CFT interpretation of the above spacetime. Since we only know the RT portion of the spacetime explicitly we will only be able to give a detailed discussion of the CFT for time $t \le t_0$. We choose a conformal frame so that the boundary metric $g_{ab}$ is given by (\ref{boundary}) for $t \le t_0$. The boundary metric (\ref{boundary}) has a special geometrical feature. The unit timelike vector 
\beq
 v^a = \left( \frac{\partial}{\partial t} \right)^a
\label{RT-old-velocity}
\eeq
is tangent to a congruence of expanding, geodesic timelike curves with vanishing rotation and shear. In fact the metric (\ref{boundary}) (without imposing any restriction on $P$) is the most general 2+1 dimensional metric admitting such a congruence.

We apply the standard AdS/CFT prescription of Ref. \cite{BalasubramanianKraus1999} to calculate the expectation value of the CFT energy-momentum tensor. The result is
\beq
\begin{array}{ll}
\displaystyle{ \langle T_{tt} \rangle = \frac{m}{4 \pi}} & \displaystyle{\langle T_{\zeta \zeta} \rangle = - \frac{1}{8\pi} \del_t \left( \frac{\del_{\zeta}^2 P}{P} \right)} \\
\displaystyle{\langle T_{t \zeta} \rangle = - \frac{1}{16 \pi} \del_{\zeta} K} & \displaystyle{ \langle T_{\zeta \zetabar} \rangle = \frac{m}{8\pi P^2}}, 
\end{array}
\label{stress-tensor-components}
\eeq
with other components related by symmetry and complex conjugation. It turns out that this can be rewritten exactly in the compact form
\beq
\langle T_{ab} \rangle = p_0 \left( 3 v_a v_b + g_{ab} \right) + \frac{1}{8\pi} Z_{(ab)},
\label{RT-stress-tensor}
\eeq
where $v_a = g_{ab} v^b = -(\d t)_a$,
\beq
p_0 = \frac{m}{8\pi},
\eeq 
and\footnote{Note the sign in the brackets: we are not projecting orthogonally to $v^a$.}
\beq
Z_{ab} = \left( \delta^d_a - v_a v^d \right) C_{dbc} v^c,
\label{Z-tensor}
\eeq
where $C_{abc}$ is the \emph{Cotton tensor} of $g_{ab}$:
\beq
C_{abc} = \nabla_c R_{ab} - \nabla_{b} R_{ac} + \frac{1}{4} \left( g_{ac} \nabla_b R - g_{ab} \nabla_c R \right).
\label{Cotton-tensor}
\eeq
In $2+1$ dimensions the Weyl tensor vanishes identically. It is the Cotton tensor that is conformally covariant and measures the deviation of the spacetime from conformal flatness.\footnote{
The Cotton tensor also played an important role in the work of Ref. \cite{Mukhopadhyay:2013gja}. Our term $Z_{(ab)}$ would vanish if $C_{abc}$ is restricted in the way discussed in that Reference.}

The above energy-momentum tensor is the sum of a conformal perfect fluid stress tensor and a three-derivative curvature term. The perfect fluid term has energy density $\rho_0=2p_0$ and constant temperature
\beq
 T_0 = \frac{3}{4\pi} (16 \pi p_0)^{1/3} = \frac{3}{4\pi} (2m)^{1/3}.
\eeq
The perfect fluid term describes a fluid that remains at rest with constant temperature in the spatially inhomogeneous, time-dependent geometry (\ref{boundary}). Such a flow does not satisfy the equation of motion of a conformal perfect fluid except in the special case for which the boundary geometry is time-independent. However, the presence of the three-derivative curvature term in (\ref{RT-stress-tensor}) ensures that the full energy-momentum tensor is conserved, $\nabla_a \langle T^{ab} \rangle = 0$, {\it provided} the function $P$ (equivalently $\lambda$) appearing in the boundary metric (\ref{boundary}) satisfies the RT equation. 

As $t\rightarrow -\infty$, $\lambda \rightarrow {\rm constant}$ so the boundary becomes conformally flat and hence the curvature term vanishes at early time. So in the far past, the energy-momentum tensor is that of a thermal state at temperature $T_0$. At finite $t$, it becomes non-thermal because of the time-dependence of the spacetime. 

The surprising feature of the result (\ref{RT-stress-tensor}) is that it depends {\it locally} on the metric. For a fixed "in" state, one would expect $\langle T_{ab} \rangle$ at a spacetime point $p$ to depend on the geometry in the entire past light cone of $p$, not on just the local geometry at $p$ \cite{Wald1977}. From an AdS/CFT perspective, this is because the boundary metric at a point $q$ in the past light cone of $p$ affects the bulk geometry near $q$ and bulk scattering leads to information from $q$ propagating to $p$. However, this does not happen here. The reason is that the algebraically special property ensures that radiation in the bulk is purely ingoing: there is no reflection back to the boundary. This is obviously non-generic, i.e., fine-tuned. This fine-tuning amounts to requiring that the function $\lambda$ in the boundary metric should satisfy the RT equation. If $\lambda$ did not satisfy this equation then $\langle T_{ab} \rangle$ would not be a local function of the metric.

%
%

%
%

\section{Conformal fluids in $2+1$ dimensions}
\label{section-conf-fluids-3-dim}

Fluid dynamics is an effective description of an interacting field theory characterized by a simple set of variables. In the simplest, uncharged case, these variables are a temperature field $T(x)$ and a velocity field $u^a(x)$ which is unit-normalized, $u_a u^a = -1$. These vary on a scale $L$ much larger than the characteristic interaction scale $L_{\mathrm{I}}$, set by the mean free path, for example. As a consequence, derivatives of $T$ and $u^a$ are increasingly smaller and fluid dynamics can be described in an expansion in derivatives of the dynamical variables.

The equations of motion of an uncharged fluid are obtained from conservation of the energy-momentum tensor
\beq
\nabla_a T^{ab} = 0.
\label{stress-tensor-conservation}
\eeq
For an uncharged fluid, $T_{ab}$ is completely determined by the $d$ degrees of freedom contained in $T$ and $u^a$. Supplementing (\ref{stress-tensor-conservation}) with an expression for $T_{ab}$ written in terms of the fluid variables thus constitutes a well-defined dynamical system, the relativistic fluid dynamical equations.

The stress tensor for a general fluid is given by
\beq
T_{ab} = \left( \rho + p \right) u_a u_b + p g_{ab} + \Pi_{ab},
\eeq
where $\rho$ is the energy density and $p$ the pressure, both of which are determined by the temperature via equations of state. The dissipative part $\Pi_{ab}$ contains the contributions constructed from derivatives of $T$ and $u^a$. Since these vary slowly, $\Pi_{ab}$ can be expanded as
\beq
\Pi_{ab} = \sum_{n \geq 1} \Pi^{(n)}_{ab},
\eeq
where $\Pi^{(n)}_{ab}$ contains $n$ derivatives of the fluid variables. As a consequence of the slow variation hypothesis, each $\Pi^{(n)}_{ab}$ is increasingly subdominant in this expansion.

Although the explicit expression of the $\Pi^{(n)}_{ab}$ can only be determined by a detailed study of the system in question, their allowed form is constrained by symmetry and other general considerations. Since we are ultimately interested in three-dimensional conformal fluids, we will restrict to this case from now on. For conformal fluids, the stress tensor must be traceless (at all orders), which imposes that
\beq
\rho = 2 p
\eeq
and that all $\Pi^{(n)}_{ab}$ must be traceless. Conformal covariance also dictates the form of the equation of state, which can be obtained by dimensional analysis:
\beq
p \propto T^3.
\eeq
Furthermore, the stress tensor must transform homogeneously under conformal transformations. Hence, only $(0,2)$ tensors with this property can appear in each $\Pi^{(n)}_{ab}$.

There is an ambiguity in the above description because $T(x)$ and $u^a(x)$ have no intrinsic definition out of equilibrium. It is usually convenient to fix this ambiguity, and the standard and natural way of doing this for uncharged fluids is by working in \emph{Landau frame}, which aligns the fluid velocity with the energy flow. More precisely, the velocity is defined to be the unique (future-directed, unit-normalized) timelike eigenvector of the stress tensor,
\beq
T_{ab} u^b = - \rho u_a,
\eeq
and the temperature is defined by identifying the corresponding eigenvalue with the energy density. This implies that the dissipative part is then transverse to $u^a$, and hence $\Pi^{(n)}_{ab} u^b = 0$ for all $n$.

In Landau frame, the form of $\Pi^{(n)}_{ab}$ is then restricted to be a linear combination of independent symmetric, traceless $(0,2)$ tensors that contain $n$ derivatives of the fluid variables, are transverse to $u^a$ and transform homogeneously under conformal transformations. By independent we mean those tensors that are not related to each other by geometric relations such as Bianchi identities or by the equations of motion (\ref{stress-tensor-conservation}). There is only a finite number of such tensors at any order, and their complete classification at first and second order in derivatives has been obtained in Ref. \cite{BaierEtAl2008} (see also \cite{Loganayagam2008, BhattacharyyaEtAl2008b, HubenyMinwallaRangamani2011}).

At first order, one finds that the equations of motion can be used to eliminate all derivatives of the temperature (equivalently the pressure) in terms of derivatives of the velocity. There is then a single contribution to the dissipative part $\Pi^{(1)}_{ab}$, namely the \emph{shear tensor} $\sigma_{ab}$ of the fluid\footnote{Here we assume a parity-invariant fluid. We will discuss parity non-invariant fluids below.}
\beq
\Pi^{(1)}_{ab} = -2\eta \, \sigma_{ab}.
\label{dissipative-1}
\eeq
The shear tensor is simply the symmetric traceless and transverse part of $\nabla_a u_b$,
\beq
\sigma_{ab} = P_{(a}{}^c P_{b)}{}^d \nabla_c u_d - \frac{P^{cd} \nabla_c u_d}{2} P_{ab},
\eeq
where
\beq
P_{ab} = g_{ab} + u_a u_b
\eeq
projects onto the subspace orthogonal to $u^a$. The transport coefficient $\eta$ is called the \emph{shear viscosity}. Its functional dependence on the temperature or, equivalently, on the pressure, is dictated by conformal covariance:
\beq
\eta = \hat{\eta} \, p^{2/3},
\eeq
where $\hat{\eta}$ is a constant. At second order, one again finds that the equations of motion allow derivatives of $p$ to be written in terms of derivatives of $u^a$. The general form of the second-order corrections in arbitrary dimensions is now well known \cite{BaierEtAl2008}, but in three dimensions they reduce to the terms written in Ref. \cite{VanRaamsdonk2008}:
\beq
\Pi^{(2)}_{ab} = 2 \tau_{\pi} \eta \, u^c \calD_c \sigma_{ab} + \lambda_2 \left( \sigma_a{}^c \omega_{cb} + \sigma_b{}^c \omega_{ca} \right),
\label{dissipative-2}
\eeq
where we are using the notation of Ref. \cite{HubenyMinwallaRangamani2011}, $\calD_a$ being the Weyl covariant derivative introduced in Ref. \cite{Loganayagam2008} (see Appendix \ref{Weyl-cov-formalism}), and $\omega_{ab}$ the \emph{rotation} or \emph{vorticity} of the fluid,
\beq
\omega_{ab} = P_{[a}{}^c P_{b]}{}^d \nabla_c u_d.
\eeq

The procedure outlined above can in principle be carried out to higher orders, but becomes increasingly complicated beyond two derivatives. Here we introduce a new formalism that involves classifying \emph{scalars} rather than tensor fields, making the task fairly simple. The new formalism is inspired by the Geroch-Held-Penrose (GHP) formalism \cite{GerochHeldPenrose1973} and the Weyl-covariant formalism of Ref. \cite{Loganayagam2008} (see Appendix \ref{Weyl-cov-formalism}). In GHP, one has two preferred null directions that one chooses as null basis vectors. In the fluid dynamical case, one has a preferred timelike congruence instead specified by the fluid velocity field (once the choice of frame, e.g. Landau frame, has been made). The remaining (spatial) basis vectors can be chosen arbitrarily and rotated at will. One is then interested in scalars that transform homogeneously under conformal transformations and spatial rotations, so that appropriate derivative operators must be defined to take this into account. In this Section we only indicate the key ideas and results, referring the reader to Appendix \ref{GHP-like-section} for more details.

We choose the fluid velocity $u^a$ to be one of the basis vectors and complete the basis with the complex-conjugate pair of vector fields $m^a, \bar{m}^a$ such that the only non-zero inner products between basis vectors are
\beq
g_{ab}u^a u^b = -1, \qquad g_{ab} m^a \bar{m}^b = 1.
\eeq
A \emph{Weyl transformation} $g_{ab} \rightarrow \Omega^2 g_{ab}$ induces a rescaling of all the basis vectors,
\beq
u^a \rightarrow \Omega^{-1} u^a, \qquad m^a \rightarrow \Omega^{-1} m^a, \qquad \mbar^a \rightarrow \Omega^{-1} \mbar^a.
\eeq
Furthermore, we can perform a rotation on $m^a,\mbar^a$,
\beq
m^a \rightarrow e^{i\lambda} m^a, \qquad \mbar^a \rightarrow e^{-i\lambda} \mbar^a,
\eeq
which we will refer to as a \emph{spin transformation}. We then project all tensor fields along this basis, so that each component thus obtained will transform in a different way under spins. We say that a scalar quantity $Q$ has definite \emph{conformal weight} $w$ and \emph{spin weight} $s$, abbreviated weight $(w,s)$ if, under the transformations above, it transforms according to
\beq
Q \rightarrow \Omega^w e^{is\lambda} Q.
\eeq
In general, however, derivatives of $Q$ will not have definite weight, even when projected along the basis. We then define new derivative operators $\Dscr, \delta, \deltabar$ which are essentially partial derivatives along each of the basis vectors $u^a,m^a,\mbar^a$ respectively, corrected by adding some ``connection'' terms to ensure that the resultant object has a definite weight. The construction of these operators and their precise definition is given in Appendix \ref{GHP-like-section}---see equations (\ref{Dscr-def}), (\ref{delta-def}), (\ref{deltabar-def}). Here we only need to point out that, if $Q$ has weight $(w,s)$ as above, then $\Dscr Q, \, \delta Q, \, \deltabar Q$ will have weights $(w-1,s)$, $(w-1,s+1)$ and $(w-1,s-1)$, respectively.

The usefulness of this formalism in fluid dynamics lies in the following. Consider a conformal fluid in 2+1 dimensions in Landau frame, i.e. the stress tensor is
\beq
T_{ab} = p(3 u_a u_b + g_{ab}) + \Pi_{ab},
\eeq
where $\Pi_{ab}$ is symmetric, traceless and transverse to $u^a$. These conditions together imply that the only non-zero components of $\Pi_{ab}$ are
\beq
\pi_2 \equiv \Pi_{ab} m^a m^b, \qquad \pi_{-2} \equiv \Pi_{ab} \mbar^a \mbar^b,
\eeq
with spins $2$ and $-2$, respectively. Reality of $T_{ab}$ implies that $\pi_{-2} = \bar{\pi}_2$, hence we need only consider the spin-2 component. In order to classify the possible contributions to $\Pi_{ab}$ in a derivative expansion, then, one only needs to find the independent scalars having spin weight 2. Thus one deals only with scalars and partial derivatives, making the task of classifying the terms at high orders much simpler. 

After projecting the relevant fluid dynamical and curvature tensor fields along the basis and using the Ricci identities (see Appendix \ref{GHP-like-section} for details), one finds only seven independent scalars, summarized in Table \ref{indep-scalars}. Apart from the pressure $p$, the fluid data comprises three scalars built from the fluid velocity: $\sigma, \bar{\sigma}$ correspond to the two independent components of the shear and $\omega$ corresponds to the single independent component of the vorticity. One can also build scalars from the curvature\footnote{It is worth emphasising that the Weyl covariant curvature tensors of Appendix \ref{Weyl-cov-formalism} involve not just the Riemann tensor of the metric, but also contributions from derivatives of the velocity.}: the $\phi_i$ in Table \ref{indep-scalars} are three particular components of the Weyl covariant Ricci tensor $\calR_{ab}$ defined in Ref. \cite{Loganayagam2008}  (see Eq. (\ref{curly-Ricci}) of Appendix \ref{Weyl-cov-formalism}).

\begin{table}[ht]

\caption{Independent scalars with definite conformal and spin weights}

\begin{center}

\begin{tabular}{c|c||c|c}

Fluid data & Weight & Curvature components & Weight \\
\hline

$p$ & $(-3,0)$ & {} & {} \\

$\sigma = m^a m^b \nabla_a u_b$ & $(-1,2)$ & $\phi_1 = u^a m^b \calR_{(ab)}$ & $(-2,1)$ \\

$\omega = i m^a \mbar^b \nabla_{[a} u_{b]}$ & $(-1,0)$ & $\phi_0 = m^a \mbar^b \calR_{(ab)}$ & $(-2,0)$ \\

$\bar{\sigma} = \mbar^a \mbar^b \nabla_a u_b$ & $(-1,-2)$ & $\bar{\phi}_1 = u^a \mbar^b \calR_{(ab)}$ & $(-2,-1)$

\end{tabular}

\end{center}

\label{indep-scalars}

\end{table}

In order to complete our formalism, we need to know how to commute derivatives and the form of the equations of motion and the Bianchi identities. If $Q$ is a scalar of weight $(w,s)$ as before, the commutators are given in terms of the fluid dynamical and curvature objects of Table \ref{indep-scalars} by
\bearr
(\Dscr \delta - \delta \Dscr) Q &=& i \omega \delta Q - \sigma \deltabar Q - 2w (\phi_1 - \deltabar \sigma + i \delta \omega) Q + s (2 \phi_1 - \deltabar \sigma + i \delta \omega) Q, \\
(\Dscr \deltabar - \deltabar \Dscr) Q &=& - i \omega \deltabar Q - \bar{\sigma} \delta Q - 2w (\bar{\phi}_1 - \delta \bar{\sigma} - i \deltabar \omega) Q - s (2 \bar{\phi}_1 - \delta \bar{\sigma} - i \deltabar \omega) Q, \\
(\delta \deltabar - \deltabar \delta) Q &=& -2i \omega \Dscr Q + (2iw \Dscr \omega - s \phi_0 )Q.
\eearr
The Bianchi identities reduce to
\bearr
0 &=& \deltabar \phi_1 - \delta \bar{\phi}_1 + \delta^2 \bar{\sigma} - \deltabar^2 \sigma - i (\Dscr^2 \omega - \delta \deltabar \omega - \deltabar \delta \omega), \\
0 &=& \Dscr \phi_0 - 2 \delta \bar{\phi}_1 - 2 \deltabar \phi_1 + \delta^2 \bar{\sigma} + \deltabar^2 \sigma + 4 \omega \Dscr \omega,
\eearr
and the fluid equations of motion $\nabla_a T^{ab} = 0$ become
\beq
2 \Dscr p + \sigma \bar{\pi}_2 + \bar{\sigma} \pi_2 = 0, \qquad \delta p + \deltabar \pi_2 = 0.
\label{eqns-motion}
\eeq
The latter imply that not all scalars quoted in Table \ref{indep-scalars} and their derivatives are independent. First, one can argue iteratively that all derivatives of $p$ can be eliminated in favour of derivatives of $u^a$ order by order using the equations of motion. More precisely, suppose that this is true to $k$th order in derivatives, so that $\pi_2$ depends on derivatives of all scalars of Table \ref{indep-scalars} except $p$. Then Eqs. (\ref{eqns-motion}) imply that, at order $k+1$, all derivatives of $p$ can be written in terms of derivatives of the other scalars and hence eliminated from $\pi_2$. 

Eqs. (\ref{eqns-motion}) also imply that the apparently one-derivative quantities $\Dscr p, \, \delta p$ are actually two-derivative quantities. Hence if we substitute $Q = p$ in the commutators above then the LHS in all cases is at least third order in derivatives. However the RHS contains terms that would \emph{a priori} be of second order in derivatives, namely $\phi_1 - \deltabar \sigma + i \delta \omega$ and $\Dscr \omega$. This means that these are in fact three-derivative quantities:
\beq
\phi_1 = \deltabar \sigma - i \delta \omega + \calO (\del^3)
\label{scalar-elim-1}
\eeq
and\footnote{For a perfect fluid, we have $\Dscr \omega = 0$ exactly. This is equivalent to the \emph{conservation of enstrophy} discussed in Ref. \cite{CarrascoEtAl2012}. $\Dscr \omega = 0$ is an equation for propagation of vorticity. Similar equations for the propagation of the shear and expansion (Raychaudhuri's equation) of $u^a$ have been used to eliminate other curvature scalars in favour of $\sigma, \bar{\sigma}, \omega$ and their derivatives, see Appendix \ref{GHP-like-section}.}
\beq
\Dscr \omega = \calO (\del^3),
\label{scalar-elim-2}
\eeq
where $\calO (\del^3)$ represents terms involving three or more derivatives. Hence, \emph{at two derivatives}, we can eliminate $\phi_1$ and its complex conjugate in terms of derivatives of $\sigma, \bar{\sigma}, \omega$ and set $\Dscr \omega = 0$. This can then be done order by order in our derivative expansion.  In particular, eliminating $\phi_1$ and $\Dscr \omega$ as in Eqs. (\ref{scalar-elim-1}), (\ref{scalar-elim-2}) in the Bianchi identities gives the single equation
\beq
\Dscr \phi_0 = \delta^2 \bar{\sigma} + \deltabar^2 \sigma + \calO (\del^4).
\label{scalar-elim-3}
\eeq
That is, $\Dscr \phi_0$ differs from derivatives of the shear by four-derivative terms and can therefore be eliminated at three derivatives. This can again be done order by order in the derivative expansion. We are thus left with only five scalars, which are summarized in Table \ref{building-blocks-deriv-exp}. Furthermore, as just explained, we can eliminate $\Dscr \omega$, $\Dscr \phi_0$ and all derivatives of $p$. 

\begin{table}[ht]

\caption{Independent objects relevant for fluid dynamics and their weights $(w,s)$}

\begin{center}

\begin{tabular}{c|c|c}

Fluid data & Curvature & Derivative operators \\
\hline

$p$: $(-3,0)$ & {} & {} \\

$\sigma$: $(-1,2)$ & {} & $\delta$: $(-1,1)$ \\

$\omega$: $(-1,0)$ & $\phi_0$: $(-2,0)$ & $\Dscr$: $(-1,0)$ \\

$\bar{\sigma}$: $(-1,-2)$ & {} & $\deltabar$: $(-1,-1)$ \\

\end{tabular}

\end{center}

\label{building-blocks-deriv-exp}

\end{table}

Classifying the various contributions to the derivative expansion in the stress tensor is now much simpler than in the usual approach. As emphasised above, we only need to consider the spin-2 component $\pi_2$. At any level in derivatives, all we need to do is classify all independent scalars built from the objects of Table \ref{building-blocks-deriv-exp} with $s = 2$. They will appear in $\pi_2$ in a linear combination with coefficients depending on $p$, and this $p$-dependence is fixed by conformal covariance. In three dimensions, $T_{ab}$ has conformal weight $w = - 1$ which implies that $\pi_2$ has $w = - 3$. 

It is also simple to classify scalars with spin $ s \ne 2$. The motivation for doing this is that we also want to define an \emph{entropy current}: a vector field $J^a$ constructed from the fluid variables whose divergence is non-negative for any flow in any background. Of course, such a vector field can be expanded in our basis,
\beq
J^a = -J_0 u^a + \bar{J}_1 m^a + J_1 \mbar^a,
\eeq
where $J_0,J_1,\bar{J}_1$ are components with spins $0$, $1$ and $-1$, respectively. 
Hence, by classifying spin-0 and spin-1 scalars, we can determine the most general form for the entropy current. The conformal weight of $J^a$ is $w = -3$ which implies that the components $J_0,J_1,\bar{J}_1$ have $w = -2$.

There is only one independent scalar involving no derivatives, namely the pressure $p$, which then determines the energy density $\rho = \rho(p) = 2p$ and the temperature $T = T(p) = \alpha p^{1/3}$ for some constant $\alpha$. 
At first order (one derivative), there are only two scalars with non-negative spin: $\sigma$ and $\omega$. The first has spin 2 and can appear in the stress tensor:
\beq
\pi_2^{(1)} = C_{\sigma} p^{2/3} \sigma,
\label{pi2-1-deriv}
\eeq
where $C_{\sigma}$ is a constant and the dependence on $p$, which is fixed by requiring that $\pi_2$ has conformal weight $w= -3$, was made explicit. Note that this agrees with equation (\ref{dissipative-1}): in our notation, the latter is rewritten as
\beq
\Pi^{(1)}_{ab} = -2\eta \, \sigma_{ab} = -2 \eta \left( \bar{\sigma} m_a m_b + \sigma \mbar_a \mbar_b \right),
\eeq
so that
\beq
\pi_2^{(1)} = \Pi^{(1)}_{ab} m^a m^b = -2 \eta \sigma.
\eeq
The shear viscosity $\eta$ is then related to $C_{\sigma}$ simply by
\beq
\label{etaCsigma}
\eta = - \frac{1}{2} C_{\sigma} p^{2/3}.
\eeq
At two derivatives, we have the following scalars:
\beq
\begin{array}{llll}
\mbox{spin 0:} & \phi_0, & \sigma \bar{\sigma}, & \omega^2 \\
\mbox{spin 1:} & \deltabar \sigma, & \delta \omega & {} \\
\mbox{spin 2:} & \Dscr \sigma, & \sigma \omega & {}
\end{array}
\label{2-deriv-scalars}
\eeq
Only those in the third line can appear in $\pi_2$:
\beq
\pi_2^{(2)} = p^{1/3} \left( C_{\Dscr \sigma} \Dscr \sigma + C_{\sigma \omega} \sigma \omega \right),
\eeq
where again the dependence on $p$ is determined by the conformal weight and the coefficients $C_Q$ are constants. This also agrees in form with the known expression (\ref{dissipative-2}) for the second-order corrections to the perfect fluid. In fact, equation (\ref{dissipative-2}) can be written in our formalism as
\beq
\Pi^{(2)}_{ab} = \left( 2 \tau_{\pi} \eta \Dscr \bar{\sigma} - 2i\lambda_2 \bar{\sigma} \omega \right) m_a m_b + \left( 2 \tau_{\pi} \eta \Dscr \sigma + 2i\lambda_2 \sigma \omega \right) \mbar_a \mbar_b,
\eeq
so that
\beq
\pi_2^{(2)} = 2 \tau_{\pi} \eta \Dscr \sigma + 2i\lambda_2 \sigma \omega
\eeq
and the coefficients in the two languages are related by
\beq
\tau_{\pi} \eta = \frac{1}{2} C_{\Dscr \sigma} p^{1/3}, \qquad \lambda_2 = - \frac{i}{2} C_{\sigma \omega} p^{1/3}.
\eeq

As it turns out, using our formalism we can easily go beyond second order and determine all independent, three-derivative scalars:
\beq
\begin{array}{lllllllll}
\mbox{spin 0:} & \delta^2 \bar{\sigma}, & \delta \deltabar \omega, & \deltabar^2 \sigma, & \bar{\sigma} \Dscr \sigma, & \sigma \Dscr \bar{\sigma}, & \omega \phi_0, & \sigma \bar{\sigma} \omega, & \omega^3 \\
\mbox{spin 1:} & \deltabar \Dscr \sigma, & \delta \phi_0, & \bar{\sigma} \delta \sigma, & \omega \deltabar \sigma, & \sigma \delta \bar{\sigma}, & \omega \delta \omega, & \sigma \deltabar \omega & {} \\
\mbox{spin 2:} & \delta^2 \omega, & \delta \deltabar \sigma, & \Dscr^2 \sigma, & \omega \Dscr \sigma, & \sigma \phi_0, & \sigma^2 \bar{\sigma}, & \sigma \omega^2 & {}
\end{array}
\label{3-deriv-scalars}
\eeq
Thus we find that there are seven independent contributions to the stress tensor at third order, in which case the transport coefficients are independent of $p$. Note that an ordering choice for derivatives has been made when writing down the scalars in (\ref{3-deriv-scalars}). For example, $\deltabar \delta \sigma$ is an equally possible spin-2 object. However, using the commutators of derivatives, we can write this in terms of $\delta \deltabar \sigma$, $\omega \Dscr \sigma$ and $\sigma \phi_0$. Hence $\deltabar \delta \sigma$ is not independent from the spin-2 scalars listed in (\ref{3-deriv-scalars}). Similar arguments hold for $\delta \deltabar \omega$ (spin 0) and $\deltabar \Dscr \sigma$ (spin 1).

We conclude this section by commenting on discrete transformations. We will call \emph{time reversal} a transformation $T$ with action
\beq
T: u^a \rightarrow - u^a,
\eeq
keeping $m^a,\mbar^a$ unchanged. In turn, a \emph{parity transformation} $P$ will act on the basis as
\beq
P: m^a \leftrightarrow \mbar^a
\eeq
with $u^a$ fixed. Note that both $T$ and $P$ change orientation, since the only independent component of the volume form $\epsilon_{abc}$ is (say) $\epsilon_{abc} u^a m^b \mbar^c$. Under time reversal, the fundamental objects of Table \ref{building-blocks-deriv-exp} change according to
\beq
T: \sigma \rightarrow - \sigma, \qquad \omega \rightarrow - \omega, \qquad \Dscr \rightarrow - \Dscr,
\eeq
with $p$, $\phi_0$ and $\delta, \deltabar$ unchanged. On the other hand, under parity, we have
\beq
P: \sigma \leftrightarrow \bar{\sigma}, \qquad \omega \rightarrow - \omega, \qquad \delta \leftrightarrow \deltabar,
\eeq
with $p$, $\phi_0$ and $\Dscr$ unchanged. We do not expect time reversal to be a symmetry of a general fluid. In fact, dissipation is precisely a feature of non-invariance under time reversal. On the other hand, we expect a large class of conformal fluids to be invariant under parity. The action of $P$ on the stress tensor
\beq
T_{ab} = 2p \left( u_a u_b + m_{(a} \mbar_{b)} \right) + \bar{\pi}_2 m_a m_b + \pi_2 \mbar_a \mbar_b
\eeq
is
\beq
P: T_{ab} \rightarrow T'_{ab} = 2p \left( u_a u_b + m_{(a} \mbar_{b)} \right) + \pi'_2 m_a m_b + \bar{\pi}'_2 \mbar_a \mbar_b.
\eeq
If the fluid is parity-invariant, i.e. $T'_{ab} = T_{ab}$, then we must have
\beq
\pi'_2 = \bar{\pi}_2.
\eeq
Any scalar $Q$ appearing in $\pi_2$ will be multiplied by a coefficient $C_Q p^{\alpha_Q}$ wich is itself invariant under parity. The scalar itself will change as $Q \rightarrow \pm \bar{Q}$, so that parity-invariance requires that $C_Q$ is real if $Q \rightarrow \bar{Q}$ and $C_Q$ is purely imaginary if $Q \rightarrow - \bar{Q}$. We can notice examples of the two behaviours above. The shear term in $\pi_2^{(1)}$ transforms as $\sigma \rightarrow \bar{\sigma}$, hence parity-invariance requires $C_{\sigma}$ or, equivalently, the shear viscosity $\eta$, to be real. On the other hand, the second term in $\pi_2^{(2)}$ changes as $\sigma \omega \rightarrow - \bar{\sigma} \omega$. Parity-invariance then requires $C_{\sigma \omega}$ to be purely imaginary, that is, $\lambda_2$ to be real.

The discussion in the previous paragraph shows that parity-violating fluids have additional transport coefficients. For example, if one does not require parity invariance, then there is another $(0,2)$ tensor that can contribute to $\Pi_{ab}$ at first order, namely \cite{SaremiSon2011}
\beq
\tilde{\sigma}_{ab} = \frac{1}{2} \left( \epsilon_{acd} u^c \sigma^d{}_b + \epsilon_{bcd} u^c \sigma^d{}_a \right).
\eeq
This has only two non-zero components in our notation,
\beq
\tilde{\sigma}_2 = \tilde{\sigma}_{ab} m^a m^b = i \sigma
\eeq
and its complex conjugate $\tilde{\sigma}_{-2} = \tilde{\sigma}_{ab} \mbar^a \mbar^b = -i \bar{\sigma}$, where we used $\epsilon_{abc} u^a m^b \mbar^c = -i$, see Appendix \ref{GHP-like-section}. If one then writes
\beq
\Pi^{(1)}_{ab} = -2 \eta \sigma_{ab} - 2 \eta_{\mathrm{H}} \tilde{\sigma}_{ab},
\eeq
one finds
\beq
\pi_2^{(1)} = \Pi^{(1)}_{ab} m^a m^b = -2 (\eta + i \eta_{\mathrm{H}}) \sigma.
\eeq
This is equivalent to having a complex coefficient $C_{\sigma}$ in (\ref{pi2-1-deriv}) and
\beq
\eta = - \frac{1}{2} \mathrm{Re} (C_{\sigma}) p^{2/3}, \qquad \eta_{\mathrm{H}} = - \frac{1}{2} \mathrm{Im} (C_{\sigma}) p^{2/3}.
\eeq
The transport coefficient $\eta_{\mathrm{H}}$ is called \emph{Hall viscosity}. Its presence introduces no dissipation \cite{SaremiSon2011}: any entropy current constructed from the fluid variables has divergence \cite{BaierEtAl2008, Romatschke2010, Bhattacharyya2012}
\beq
\nabla_a J^a = - \frac{2}{3} \frac{s}{p^{1/3}} \mathrm{Re} (C_{\sigma}) \sigma \bar{\sigma} + \calO (\del^3),
\label{div-entropy}
\eeq
where $s$ is the entropy density. Some examples can be found in Refs. \cite{SaremiSon2011,JensenEtAl2011,ChenEtAl2012}, which recently investigated holographic models dual to parity-violating fluids in 2+1 dimensions exhibiting a non-zero Hall viscosity and other analogous transport coefficients.

%
%

\section{Fluid/gravity interpretation of RT}
\label{section-RT-fluid-gravity}

\subsection{Introduction}

We determined above the expectation value of the CFT energy-momentum tensor in a state dual to a RT solution. As emphasized above, the result (\ref{RT-stress-tensor}) is exact, it does not assume any derivative expansion. But now let us consider the case in which the background geometry (\ref{boundary}) is slowly varying compared to the scale set by the inverse temperature of the fluid. Specifically, we assume that the background geometry varies spatially over a length scale $L$, so the Gaussian curvature of $g_{(2)}$ is $K= {\cal O}(L^{-2})$. The RT equation then implies that temporal variations in the background geometry occur over the time scale $L^4$ so the time variation is very slow compared to the scale of the spatial variation. 

We can use the RT equation to eliminate time derivatives from our CFT energy-momentum tensor, and then expand it according to the number of \emph{spatial} derivatives. It is clear that this will give a leading order perfect fluid piece of the form discussed above and corrections involving three or more spatial derivatives. 

We want to compare this with the known results for the derivative expansion of the energy-momentum tensor dual to a general fluid flow in a general background, as given by the fluid/gravity correspondence \cite{BhattacharyyaEtAl2008a, BhattacharyyaEtAl2008b}. In order to do so, we wish to employ our new formalism developed above. We then start by defining the vector fields
\beq
\left( m^0 \right)^a = P \left( \frac{\del}{\del \zetabar} \right)^a, \qquad \left( \mbar^0 \right)^a = P \left( \frac{\del}{\del \zeta} \right)^a.
\eeq
Together with $v^a$ of Eq. (\ref{RT-old-velocity}), these form a basis for the tangent space at every point and satisfy the conditions required for our formalism:
\beq
g_{ab} v^a v^b = -1, \qquad g_{ab} \left( m^0 \right)^a \left( \mbar^0 \right)^b = 1,
\eeq
with all other inner products zero. But in order to compare our results with previous results, we must ensure that we are comparing like with like. Our result (\ref{RT-stress-tensor}) is {\it not} in Landau frame so we need to perform a field redefinition to convert to Landau frame. This amounts to solving the eigenvalue problem
\beq
 T_{ab} u^b = -\rho u_a,
\label{Landau-frame-def}
\eeq
which defines a new energy density $\rho$ and a new velocity field $u^a$. The Landau frame velocity differs from $v^a$ used above by terms involving three or more spatial derivatives. Since $v^a$ has vanishing rotation and shear, it follows that $u^a$ has rotation and shear involving four or more spatial derivatives (in fact we will see that the rotation vanishes even in Landau frame). This should be contrasted with a generic flow, for which rotation and shear are really one-derivative quantities. Hence, for this particular flow, the corrections to the perfect fluid written explicitly in (\ref{dissipative-1}) and (\ref{dissipative-2}) involve four or more spatial derivatives so they are subleading compared to possible three-derivative terms in $\Pi_{ab}$.

Since we know the stress tensor (\ref{RT-stress-tensor}) explicitly, we can convert to Landau frame and determine explicitly the corrections to the perfect fluid to the desired order. Using our formalism above, we can then classify all scalars that might appear in these corrections and compare with our known result. This allows us to constrain some transport coefficients at higher order.

\subsection{Landau frame results}
\label{RT-fluid-gravity}

In its full generality, the eigenvalue problem (\ref{Landau-frame-def}) cannot be solved exactly in a useful way, so we proceed to solve it in a \emph{spatial} derivative expansion. As explained above, time variations are much slower than spatial variations. We use the RT equation to convert time derivatives into spatial derivatives and write
\beq
\rho = \sum_{k \geq 0} \rho^{(k)}, \qquad u_a = \sum_{k \geq 0} u^{(k)}_a,
\eeq
where $\rho^{(k)}$ and $u^{(k)}_a$ are each supposed to contain $k$ spatial derivatives of $P$. We note that, in the coordinate system used above, $\langle T_{ab} \rangle$ given in (\ref{stress-tensor-components}) has components involving no derivatives, $\langle T_{tt} \rangle$, $\langle T_{\zeta \zetabar} \rangle$; three derivatives, $\langle T_{t \zeta} \rangle$, $\langle T_{t \zetabar} \rangle$; and six (spatial) derivatives, $\langle T_{\zeta \zeta} \rangle$, $\langle T_{\zetabar \zetabar} \rangle$. This implies that the energy density and velocity will have corrections only for those values of $k$ which are multiples of 3. Up to six spatial derivatives, we find
\bearr
\rho = 2p &=& \rho^{(0)} + \rho^{(6)} + \calO (L^{-9}) \nonumber \\ 
          &=& \frac{m}{4\pi} - \frac{P^2}{48\pi m} \del_{\zeta} K \, \del_{\zetabar} K + \calO (L^{-9})
\eearr
and
\bearr
u^a &=& \left( u^{(0)} \right)^a + \left( u^{(3)} \right)^a + \left( u^{(6)} \right)^a + \calO (L^{-9}) \nonumber \\
    &=& v^a + \frac{P}{6m} \left[ \del_{\zeta} K  \left( m^0 \right)^a + \del_{\zetabar} K  \left( \mbar^0 \right)^a \right] + \left( \frac{P}{6m} \right)^2 \del_{\zeta} K \, \del_{\zetabar} K \, v^a + \calO (L^{-9}). 
\eearr
The proportionality factor $\alpha$ in $T = \alpha p^{1/3}$ is determined by the fluid/gravity map \cite{BhattacharyyaEtAl2008b}:
\beq
T = \frac{3}{4\pi} (16 \pi p)^{1/3}.
\eeq
We likewise need to correct the vectors $\left( m^0 \right)^a, \left( \mbar^0 \right)^a$. Up to six spatial derivatives, we can choose
\bearr
m^a &=& \left( m^{(0)} \right)^a + \left( m^{(3)} \right)^a + \left( m^{(6)} \right)^a + \calO (L^{-9}) \nonumber \\
    &=& \left( m^0 \right)^a + \frac{P}{6m} \del_{\zetabar} K  v^a + \frac{1}{2} \left( \frac{P}{6m} \right)^2 \left[ \del_{\zeta} K \del_{\zetabar} K \left( m^0 \right)^a + \left( \del_{\zetabar} K \right)^2 \left( \mbar^0 \right)^a \right] + \calO (L^{-9}), \nonumber \\
		&\quad& {}
\eearr
with $\mbar^a$ determined by complex conjugation.

Using these definitions and working always till six spatial derivatives, using the RT equation to eliminate time derivatives, it is now a fairly simple matter to determine the relevant scalars that we need. We find that this flow has a non-zero shear,
\beq
\sigma = \frac{P^2}{6m} \left( \del^2_{\zetabar} K + 2 \del_{\zetabar} \ln P \del_{\zetabar} K \right) + \calO (L^{-7}),
\eeq
but vanishing rotation, $\omega = \calO (L^{-7})$. As anticipated, the shear involves four spatial derivatives for this particular flow. The remaining connection coefficients (see Appendix \ref{GHP-like-section}) required to define the derivative operators are found to be
\bearr
a      &=& \calO (L^{-7}), \\
\theta &=& \calO (L^{-7}), \\
\tau   &=& i \frac{P^2}{6m} \left( \del_{\zeta} \ln P \, \del_{\zetabar} K - \del_{\zetabar} \ln P \, \del_{\zeta} K \right) + \calO (L^{-7}), \\
\kappa &=& - P \del_{\zetabar} \ln P + \calO (L^{-7}).
\eearr
One can then show that $A_a = \calO (L^{-7})$, so that the Weyl-covariant Ricci tensor $\calR_{ab}$ of Appendix \ref{Weyl-cov-formalism}, Eq. (\ref{curly-Ricci}), differs from the usual Ricci tensor $R_{ab}$ by an eight-derivative term, $\calR_{ab} = R_{ab} + \calO (L^{-8})$. The curvature scalar $\phi_0$ is then given simply by
\beq
\phi_0 = K + \calO (L^{-8}).
\label{phi0-RT}
\eeq

We can now classify all scalars built solely from $\sigma$, $\bar{\sigma}$, $\phi_0$ and their derivatives and containing no more than six spatial derivatives. A partial classification in the general case was given in Section \ref{section-conf-fluids-3-dim}, Eqs. (\ref{pi2-1-deriv}), (\ref{2-deriv-scalars}), (\ref{3-deriv-scalars}). The only one-derivative scalar in the general classification that is relevant here is the shear $\sigma$, which we know is in fact a four spatial derivative object for this particular flow. For a general flow, the scalars $\sigma \bar{\sigma}$ and $\Dscr \sigma$ would be two-derivative quantities. However, in our case, they both contain at least eight spatial derivatives, i.e. are both $\calO (L^{-8})$, and hence will be discarded. At two derivatives, we are thus left with
\beq
\mbox{2 derivatives:} \qquad \left\{
\begin{array}{ll}
\mbox{spin 0:} & \phi_0 \\
\mbox{spin 1:} & \deltabar \sigma
\end{array} \right.
\label{2-deriv-scalars-RT}
\eeq
Again, these are two-derivative quantities for a general flow and are thus labelled as such here. But in our case, although $\phi_0$ does contain a two spatial derivative contribution, Eq. (\ref{phi0-RT}), $\deltabar \sigma$ turns out to be a five spatial derivative object. Similarly, at three derivatives we discard all terms with contributions smaller than $\calO (L^{-6})$. It is easy to see that we are then left with
\beq
\mbox{3 derivatives:} \qquad \left\{
\begin{array}{lll}
\mbox{spin 0:} & \delta^2 \bar{\sigma}, & \deltabar^2 \sigma \\
\mbox{spin 1:} & \delta \phi_0 & {} \\
\mbox{spin 2:} & \delta \deltabar \sigma, & \sigma \phi_0 
\end{array} \right.
\label{3-deriv-scalars-RT}
\eeq

As $\sigma = \calO (L^{-4})$, there are likely to be four-derivative scalars giving a contribution to the stress tensor which is comparable to (\ref{pi2-1-deriv}). For the same reason, only those four-derivative scalars built solely from the curvature scalar $\phi_0$ will have a contribution that is $\calO (L^{-4})$. It is then easy to find such quantities:
\beq
\mbox{4 derivatives (curvature):} \qquad \left\{
\begin{array}{lll}
\mbox{spin 0:} & \phi_0^2, & \delta \deltabar \phi_0 \\
\mbox{spin 2:} & \delta^2 \phi_0
\end{array} \right.
\label{4-deriv-scalars-RT}
\eeq
In fact, one finds that
\beq
\delta^2 \phi_0 = 6m \sigma + \calO (L^{-7}).
\eeq
To order $\calO (L^{-6})$, then, there is a degeneracy between derivatives of curvature and shear. This plays an important role in the RT case because the two contributions must cancel in the stress tensor, as we observe no four spatial derivative contributions in our exact expression. We will see below that this allows us to determine uniquely the transport coefficient associated with $\delta^2 \phi_0$. On the other hand, at higher orders some contributions arising from curvature and shear mix in such a way that we can only constrain some linear combinations of the corresponding transport coefficients.

In any case, a similar reasoning applies to higher orders and we can classify the curvature terms having five and six spatial derivatives:
\beq
\mbox{5 derivatives (curvature):} \qquad \phi_0 \delta \phi_0, \quad \deltabar \delta^2 \phi_0 \qquad \mbox{(spin 1)}
\label{5-deriv-scalars-RT}
\eeq
\beq
\mbox{6 derivatives (curvature):} \qquad \left\{
\begin{array}{lllll}
\mbox{spin 0:} & \phi_0^3, & \phi_0 \delta \deltabar \phi_0, & \delta \phi_0 \, \deltabar \phi_0, & \delta^2 \deltabar^2 \phi_0 \\
\mbox{spin 2:} & \phi_0 \delta^2 \phi_0, & (\delta \phi_0)^2, & \delta \deltabar \delta^2 \phi_0 & {}
\end{array} \right.
\label{6-deriv-scalars-RT}
\eeq
Now we write down the most general combination of the above scalars that can contribute to $\pi_2$ up to six spatial derivatives:
\bearr
\pi_2 &=& C_{\sigma} p^{2/3} \sigma + C_{\sigma \phi_0} \sigma \phi_0 + C_{\delta \deltabar \sigma} \delta \deltabar \sigma + C_{\delta^2 \phi_0} p^{-1/3} \delta^2 \phi_0 \nonumber \\
      &\quad& + p^{-1} \left[ C_{\phi_0 \delta^2 \phi_0} \phi_0 \delta^2 \phi_0 + C_{(\delta \phi_0)^2} (\delta \phi_0)^2 + C_{\delta \deltabar \delta^2 \phi_0} \delta \deltabar \delta^2 \phi_0 \right],
\eearr
where the $C_Q$ are constants. An explicit computation now gives
\bearr
\pi_2 &=& \frac{P^2}{6m} p_0^{2/3} \left( C_{\sigma} + 48\pi C_{\delta^2 \phi_0} \right) \left( \del^2_{\zetabar} K + 2 \del_{\zetabar} \ln P \, \del_{\zetabar} K \right) \nonumber \\
      &\quad& + \frac{P^2}{6m} \left( C_{\sigma \phi_0} + 48\pi C_{\phi_0 \delta^2 \phi_0} + C_{\delta \deltabar \sigma} + 48\pi C_{\delta \deltabar \delta^2 \phi_0} \right) K \left( \del^2_{\zetabar} K + 2 \del_{\zetabar} \ln P \, \del_{\zetabar} K \right) \nonumber \\
			&\quad& + \frac{P^2}{12m} \left( C_{\delta \deltabar \sigma} + 48\pi C_{\delta \deltabar \delta^2 \phi_0} \right) \left( \del^2_{\zetabar} \Delta K + 2 \del_{\zetabar} \ln P \, \del_{\zetabar} \Delta K \right) \nonumber \\
			&\quad& + \frac{P^2}{6m} \left( C_{\delta \deltabar \sigma} + 48\pi C_{\delta \deltabar \delta^2 \phi_0} + 48\pi C_{(\delta \phi_0)^2} \right) (\del_{\zetabar}K)^2 + \calO (L^{-7}). \label{pi2-1}
\eearr
We want to compare this expression with the one obtained from the exact stress tensor (\ref{RT-stress-tensor}). This is given up to six spatial derivatives by
\bearr
\pi_2 &=& \langle T_{ab} \rangle m^a m^b \nonumber \\
      &=& - \frac{P^2}{96\pi m} \left[ \del^2_{\zetabar} \Delta K + 2 \del_{\zetabar} \ln P \, \del_{\zetabar} \Delta K + (\del_{\zetabar}K)^2 \right] + \calO (L^{-9}). \label{pi2-2}
\eearr
We immediately see that the latter expression contains no four spatial derivative terms. Therefore, their coefficient must cancel in (\ref{pi2-1}) and we obtain
\beq
C_{\delta^2 \phi_0} = - \frac{1}{48\pi} C_{\sigma}.
\eeq
This is a fluid dual to Einstein gravity, and hence is parity-invariant. The coefficient $C_{\sigma}$ is related to the shear viscosity by (\ref{etaCsigma}) and the latter is known from the fluid/gravity correspondence \cite{VanRaamsdonk2008}: 
\beq
\eta  = \frac{1}{16\pi} \left( \frac{4\pi T}{3} \right)^2,
\eeq
giving
\beq
C_{\sigma} = - \frac{2}{(16 \pi)^{1/3}},
\eeq
and hence
\beq
C_{\delta^2 \phi_0} = \frac{2}{3 (16 \pi)^{4/3}}.
\label{4-deriv-transport-coeff-RT}
\eeq
By comparing (\ref{pi2-1}) and (\ref{pi2-2}), we are furthermore able to determine a sixth-order transport coefficient explicitly,
\beq
C_{(\delta \phi_0)^2} = \frac{1}{768 \pi^2},
\eeq
and constrain the other four according to
\beq
C_{\sigma \phi_0} + 48\pi C_{\phi_0 \delta^2 \phi_0} = - C_{\delta \deltabar \sigma} - 48\pi C_{\delta \deltabar \delta^2 \phi_0} = \frac{1}{8\pi}.
\eeq

\subsection{Entropy production}
\label{RT-entropy-current}

In the fluid description we wish to associate an entropy current to the fluid, i.e., a vector $J^a$ constructed from the fluid temperature and velocity as well as the background curvature, such that $J^a$ reduces to the usual result for a fluid in equilibrium and such that $\nabla_a J^a \ge 0$ for all flows and backgrounds. As has been discussed e.g. in Refs. \cite{Romatschke2010, Bhattacharyya2012}, there is not a unique definition of $J^a$: several different choices might have the desired property.

For a general flow of a conformal fluid, the leading order result for $\nabla_a J^a$ in a derivative expansion is independent of the ambiguity in defining $J^a$, Eq. (\ref{div-entropy}). In general, $\sigma \bar{\sigma}$ appearing in (\ref{div-entropy}) is a two-derivative term. However, for our particular flow, $\sigma$ is a four spatial derivative term and hence this is an eight spatial derivative contribution. Hence it is conceivable that there are terms in (\ref{div-entropy}) which are subleading for a general flow but nevertheless determine the leading order behaviour for our solution. For example, there might be a term on the RHS of (\ref{div-entropy}) proportional to $\phi_0^2$ (four derivatives). However, Ref. \cite{Bhattacharyya2012} has shown that such four-derivative pure curvature terms cannot arise on the RHS of (\ref{div-entropy}) (see also \cite{Romatschke2010}). Nevertheless, higher-order contributions such as $\delta \phi_0 \deltabar \phi_0$ (six derivatives) and $\phi_0^4$ (eight derivatives) are not excluded, and for our flow these would be more important, or as important, as  $\sigma \bar{\sigma}$ in (\ref{div-entropy}).

Ref. \cite{BhattacharyyaEtAl2008c} showed how to construct an entropy current with the desired property by using null geodesics to define a map from the conformal boundary to the event horizon, then using this map to pull back the volume form on a cross-section of the horizon to give a corresponding form on the boundary. Dualizing this gives the entropy current. Positivity of $\nabla_a J^a$ then follows from the Hawking area theorem in the bulk. As discussed in Ref. \cite{BhattacharyyaEtAl2008c}, there is ambiguity present in this construction since one can make different choices for the null geodesics used to construct the map from boundary to event horizon. However, the simplest choice is to use the ingoing null geodesics used in the construction of the bulk solution. In our case, we will use the preferred set of ingoing null geodesics defined by the algebraically special nature of the bulk.

The first step is to determine the location of the event horizon in the bulk. Here it is important that we know that our RT spacetime is part of an extended RT spacetime which settles down to Schwarzschild-AdS in the future, and such that the whole spacetime is slowly varying. Only then can we be sure that the location of the event horizon can be obtained using a derivative expansion. At leading order in the derivative expansion, corresponding to a planar Schwarzschild-AdS solution, the event horizon is located at $r=r_+$ where
\beq
 r_+ = (2 m)^{1/3}
\eeq
We now write the location of the event horizon as $r=r_+ + f(t,x^i)$ where $x^i$ are coordinates on ${}^{(2)}M$. A 1-form normal to the horizon is $\mathbf{n} = \d r - \d f$. The condition that this be null gives
\beq
 0= \left[  K + \frac{1}{6m} r \Delta K + F(r) - 2 \del_t f + \frac{1}{r^2} \left(\hat{ \nabla} f \right)^2 \right]_{r=r_++f}
\label{nullcond}
\eeq
where $F(r) = (r^2/l^2)(1-r_+^3/r^3)$, $\hat{\nabla}$ is the Levi-Civita connection associated to the metric $g_{(2)}$ on ${}^{(2)}M$ and $(\hat{\nabla} f)^2$ is calculated using $g_{(2)}$. We now assume that $f$ can be expanded in derivatives:
\beq
 f = f_1 + f_2 + f_3 + \ldots
\eeq
where $f_n$ is a quantity involving $n$ spatial derivatives. Substituting into (\ref{nullcond}) and solving order by order one finds that $f_n=0$ for odd $n$ and
\bearr
 f_2 &=&  -\frac{r_+^2}{6m} K, \\
 f_4 &=& -\frac{r_+^3}{(6m)^2} \Delta K, \\
 f_6 &=& \frac{r_+^4}{(6m)^3} \left[ \frac{1}{3} K^3 - K \Delta K - \Delta \Delta K - \left( \hat{\nabla} K \right)^2 \right].
\eearr
Now we need to calculate the volume form on a cross-section of the event horizon. Let $\ell^a$ be a null vector field transverse such that $\ell^a n_a =-1$ on the horizon and $\eta_{abcd}$ the spacetime volume form. Then the volume form on a horizon cross-section is
\beq
 X_{ab} = \eta_{abcd} \ell^c n^d
\eeq
We can choose $\ell^a = - \partial/\partial r$. A calculation then gives
\beq
 \mathbf{X}  = \frac{1}{2} (r_++f)^2 \sqrt{g_{(2)}} \epsilon_{ij} \d x^i \wedge \d x^j - \d t \wedge \sqrt{g_{(2)}} \epsilon_{ij} (\hat{\nabla} f)^j \d x^i
\label{volform}
\eeq
where $g_{(2)}$ is used to raise indices in the final term. We can define a map which sends a boundary point $(t,x^i)$ to the point $(t,r_++f(t,x^i),x^i)$ on the event horizon. Pulling back the above 2-form using this map gives a 2-form $\mathbf{X}$ on the boundary whose expression is identical to (\ref{volform}). Finally, we want to associate the (pull-back of) the horizon area form with a notion of entropy. In classical black hole thermodynamics, the entropy of a black hole is related to its area by $S = A/4$. We then dualize the above 2-form using the boundary metric to obtain an entropy \emph{current}
\beq
\mathbf{J} = \frac{1}{4} \left( \ast \mathbf{X} \right),
\eeq
with the appropriate factor of $1/4$ relating entropy and area. This gives
\beq 
J^a = \frac{(r_++f)^2}{4} \left( \frac{\partial}{\partial t} \right)^a - \frac{1}{4} g_{(2)}^{ij} \partial_j f \left( \frac{\partial}{\partial x^i} \right)^a.
\label{RT-entropy-exact}
\eeq
This is an exact result, no derivative expansion has been assumed here. We can, however, expand this result in spatial derivatives and compare with the general expression obtained in the fluid/gravity map. The latter is given to second order in derivatives by \cite{BhattacharyyaEtAl2008b}
\beq
J^a = su^a + \frac{1}{4} u^a \left( A_1 \, \sigma_{bc} \sigma^{bc} + A_2 \, \omega_{bc} \omega^{bc} + A_3 \calR \right) + \frac{1}{4} \left( B_1 \, \calD_b \sigma^{ba} + B_2 \, \calD_b \omega^{ba} \right) + \ldots,
\label{entropy-gen}
\eeq
where
\beq
s = \frac{1}{4} \left( \frac{4\pi T}{3} \right)^{2} = \frac{(16\pi p)^{2/3}}{4}
\label{entropy-density}
\eeq
is the entropy density, and the coefficients $A_i,B_i$ are constants, which are fixed by the fluid/gravity map \cite{BhattacharyyaEtAl2008b}. In our formalism, equation (\ref{entropy-gen}) becomes
\bearr
J^a &=& \frac{1}{4} \left[ (16\pi p)^{2/3} + (2A_1 + 2A_3 + 2B_1) \sigma \bar{\sigma} + (2A_2 - 2A_3 + 2B_2) \omega^2 + 2A_3 \phi_0 \right] u^a \nonumber \\
    &\quad& + \frac{1}{4} (B_1 \delta \bar{\sigma} - iB_2 \deltabar \omega) m^a + \frac{1}{4} (B_1 \deltabar \sigma + iB_2 \delta \omega) \mbar^a + \ldots. \label{entropy-new-form}
\eearr
In the RT case, $\sigma \bar{\sigma}$ and $\delta \omega$ contain at least eight spatial derivatives, and the $\omega^2$ term is even more negligible. Therefore, the only relevant constants in our case are $A_3$ and $B_1$, which are fixed by the fluid/gravity map as \cite{BhattacharyyaEtAl2008b}
\beq
B_1 = -2A_3 = \frac{2}{3}.
\label{B1A3}
\eeq

However, similarly to our analysis of the energy-momentum tensor, Eq. (\ref{entropy-new-form}) will not give the leading order expression in the RT case. For example, the leading order contribution of $\deltabar \sigma$, which is a two-derivative object in general, involves five spatial derivatives in our case. But there are other terms in (\ref{entropy-new-form}) which would be subleading for a generic flow, namely three- and four-derivative contributions, and nevertheless are important here. These terms are built from the curvature and were already determined in Section \ref{RT-fluid-gravity} above, namely $\delta \phi_0$, $\phi_0^2$ and $\delta \deltabar \phi_0$---see Eqs. (\ref{3-deriv-scalars-RT}), (\ref{4-deriv-scalars-RT}). In fact, with the aid of our new formalism, we can go to higher orders by considering the contributions involving up to six spatial derivatives. We begin by expanding our exact result in our usual coordinate system. To six spatial derivatives, we find
\bearr
J_0 &=& - u_a J^a \nonumber \\
    &=& - \frac{r_+^2}{4} + \frac{1}{6} K + \frac{1}{18 r_+^2} \left( \Delta K - \frac{1}{2} K^2 \right) \nonumber \\
		&\quad& + \frac{1}{54 r_+^4} \left( \Delta \Delta K - \frac{1}{3} K^3 + \frac{7}{2} P^2 \del_{\zeta} K \, \del_{\zetabar} K \right) + \calO (L^{-7})
\eearr
and
\bearr
J_1 &=& m_a J^a \nonumber \\
    &=& \frac{P}{36 r_+^3} \left( \del_{\zetabar} \Delta K + 2 K \del_{\zetabar} K \right) + \calO (L^{-7}).
\eearr
On the other hand, we have determined in Section \ref{RT-fluid-gravity} above all spin-0 and spin-1 scalars involving no more than six spatial derivatives, Eqs. (\ref{2-deriv-scalars-RT}), (\ref{3-deriv-scalars-RT}), (\ref{4-deriv-scalars-RT}), (\ref{5-deriv-scalars-RT}), (\ref{6-deriv-scalars-RT}). We can use them to write down the most general expression for $J_0$ and $J_1$ containing up to six spatial derivatives and compare them with the previous equations. For $J_0$, we have
\bearr
J_0 &=& A_0 p^{2/3} + B_{\phi_0} \phi_0 + p^{-1/3} \left( B_{\deltabar^2 \sigma} \deltabar^2 \sigma + B_{\delta^2 \bar{\sigma}} \delta^2 \bar{\sigma} \right) + p^{-2/3} \left( B_{\phi_0^2} \phi_0^2 + B_{\delta \deltabar \phi_0} \delta \deltabar \phi_0 \right) \nonumber \\
    &\quad& + p^{-4/3} \left( B_{\phi_0^3} \phi_0^3 + B_{\phi_0 \delta \deltabar \phi_0} \phi_0 \delta \deltabar \phi_0 + B_{\delta \phi_0 \, \deltabar \phi_0} \delta \phi_0 \, \deltabar \phi_0 + B_{\delta^2 \deltabar^2 \phi_0} \delta^2 \deltabar^2 \phi_0 \right),
\eearr
where $A_0$ and the $B_Q$ are constants, and the dependence on $p$ for each term is fixed by imposing the conformal weight of $J_0$ to be $w = -2$. Similarly, for $J_1$ we write
\beq
J_1 = B_{\deltabar \sigma} \deltabar \sigma + p^{-1/3} B_{\delta \phi_0} \delta \phi_0 + p^{-1} \left( B_{\phi_0 \delta \phi_0} \phi_0 \delta \phi_0 + B_{\deltabar \delta^2 \phi_0} \deltabar \delta^2 \phi_0 \right),
\eeq
where the conformal weight is also $w = -2$. Note that, in these two equations, the various contributions are ordered according to their number of derivatives in a \emph{general} derivative expansion. An explicit computation then gives
\bearr
J_0 &=& \frac{A_0 r_+^2}{(16 \pi)^{2/3}} + B_{\phi_0} K + \frac{(16 \pi)^{2/3}}{2 r_+^2} \left( B_{\delta \deltabar \phi_0} \Delta K + 2 B_{\phi_0^2} K^2 \right) \nonumber \\
    &\quad& + \frac{(16 \pi)^{1/3}}{12 r_+^4} \left( B_{\deltabar^2 \sigma} + B_{\delta^2 \bar{\sigma}} + 48 \pi B_{\delta^2 \deltabar^2 \phi_0} \right) \Delta \Delta K \nonumber \\
    &\quad& + \frac{(16 \pi)^{1/3}}{6 r_+^4} \left[  B_{\deltabar^2 \sigma} + B_{\delta^2 \bar{\sigma}} + 48 \pi \left( B_{\phi_0 \delta \deltabar \phi_0} + B_{\delta^2 \deltabar^2 \phi_0} \right) \right] K \Delta K + \frac{(16 \pi)^{4/3} B_{\phi_0^3}}{r_+^4} K^3 \nonumber \\
    &\quad& + \frac{(16 \pi)^{1/3}}{3 r_+^4} \left[ - \frac{A_0}{24 \pi} + B_{\deltabar^2 \sigma} + B_{\delta^2 \bar{\sigma}} + 48 \pi \left( B_{\delta \phi_0 \, \deltabar \phi_0} + B_{\delta^2 \deltabar^2 \phi_0} \right) \right] P^2 \del_{\zeta} K \, \del_{\zetabar} K + \calO (L^{-7}) \nonumber \\
    &\quad& {}
\eearr
and
\bearr
J_1 &=& \frac{(16 \pi)^{1/3}}{r_+} P B_{\delta \phi_0} \del_{\zetabar} K + \frac{P}{6 r_+^3} \left( B_{\deltabar \sigma} + 48 \pi B_{\deltabar \delta^2 \phi_0} \right) \del_{\zetabar} \Delta K \nonumber \\
    &\quad& + \frac{P}{3 r_+^3} \left[ B_{\deltabar \sigma} + 48 \pi \left( B_{\phi_0 \delta \phi_0} + B_{\deltabar \delta^2 \phi_0} \right) \right] K \del_{\zetabar} K + \calO (L^{-7}),
\eearr
where we have now expanded in \emph{spatial} derivatives. We can now compare these with the results obtained from expanding the RT entropy current in derivatives to determine the unknown coefficients. We first find
\beq
A_0 = - \frac{(16 \pi)^{2/3}}{4},
\eeq
which is consistent with the association
\beq
s = - A_0 p^{2/3},
\eeq
see Eq. (\ref{entropy-density}). Next we determine uniquely
\beq
B_{\phi_0} = \frac{1}{6}.
\eeq
This coefficient is related to $A_3$ of Ref. \cite{BhattacharyyaEtAl2008b} by
\beq
B_{\phi_0} = - \frac{A_3}{2},
\eeq
so our results are compatible with the fluid/gravity prediction for $A_3$, Eq. (\ref{B1A3}). We are furthermore able to determine uniquely the constants
\beq
B_{\delta \deltabar \phi_0} = -4 B_{\phi_0^2} = \frac{1}{9 (16 \pi)^{2/3}},
\eeq
\beq
B_{\phi_0 \delta \deltabar \phi_0} = 12 B_{\phi_0^3} = \frac{8}{7} B_{\delta \phi_0 \, \deltabar \phi_0} = - \frac{2}{27 \pi (16 \pi)^{4/3}}
\eeq
and
\beq
B_{\delta \phi_0} = B_{\phi_0 \delta \phi_0} = 0,
\eeq
but the remaining coefficients can only be constrained in the following linear combinations:
\beq
B_{\deltabar^2 \sigma} + B_{\delta^2 \bar{\sigma}} + 48 \pi B_{\delta^2 \deltabar^2 \phi_0} = - 48 \pi B_{\phi_0 \delta \deltabar \phi_0} = \frac{2}{9 (16\pi)^{1/3}},
\eeq
\beq
B_{\deltabar \sigma} + 48 \pi B_{\deltabar \delta^2 \phi_0} = \frac{1}{6}.
\eeq
Our coefficient $B_{\deltabar \sigma}$ is related to $B_1$ of Ref. \cite{BhattacharyyaEtAl2008b} by
\beq
B_{\deltabar \sigma} = \frac{B_1}{4}.
\eeq
Our results only allowed us to determine a linear combination of $B_{\deltabar \sigma}$ and $B_{\deltabar \delta^2 \phi_0}$. If we then use $B_1 = -2A_3$ from the fluid/gravity map, we can determine 
\beq
B_{\deltabar \sigma} = B_{\phi_0} = \frac{1}{6} 
\eeq
and then
\beq
B_{\deltabar \delta^2 \phi_0} = 0.
\eeq
The first of these is in agreement with the known value for $B_1$, Eq. (\ref{B1A3}). In fact, it is possible to show in general that non-negativity of the divergence of the entropy current requires $B_{\deltabar \sigma} = B_{\phi_0}$ or, equivalently, $B_1 = -2A_3$ \cite{Romatschke2010, Bhattacharyya2012}.

We now go back to our exact expression (\ref{RT-entropy-exact}) for the entropy current and calculate its divergence. Using the RT equation we obtain
\beq
 \nabla_a J^a = -\frac{1}{24m} (r_++f)^2 \Delta K + \frac{1}{2} (r_++f) \partial_t f - \frac{1}{4} \Delta f.
\eeq
We can now substitute the derivative expansion for $f$ into the RHS. The RHS involves four (spatial) derivative terms but it is easy to see that these cancel. There are no terms involving an odd number of derivatives. With some work (using the RT equation to eliminate time derivatives) one finds that the six spatial derivative terms also cancel. The first non-vanishing contribution arises at eight spatial derivatives:
\beq
  \nabla_a J^a = \frac{r_+^4}{2(6m)^3} \left[ \hat{\nabla}_i \hat{\nabla}_j K - \frac{1}{2} (g_{(2)})_{ij} \Delta K \right]^2 + \calO (L^{-10}).
\eeq
Comparing this with the expression for the shear of the fluid determined above, we find that this agrees precisely with the shear squared term in (\ref{div-entropy}). 

This result is perhaps surprising: for this particular fluid flow, the shear squared term is an eight spatial derivative term. But one might have expected to see terms on the RHS above with fewer spatial derivatives, constructed from the curvature of the background metric. The results of Ref. \cite{Bhattacharyya2012} already rule out the possibility of a term of the form $\phi_0^2$, but one might have expected to find terms such as $\delta \phi_0 \deltabar \phi_0$ (six derivatives) and $\phi_0^4$ (eight derivatives). Our result shows that such terms are absent up to (and including) eight derivatives.

%
%

\section{The Kerr-AdS solution}
\label{Kerr-AdS-section}


The Kerr-AdS metric is given in Boyer-Lindquist coordinates and in units such that $\Lambda = -3$ by \cite{GriffithsPodolsky2012}
\beq
ds^2 = - \frac{\Delta_r}{\Xi^2 \rho^2} (\d t - a \sin^2 \theta \, \d \phi)^2 + \frac{\rho^2}{\Delta_r} \d r^2 + \frac{\rho^2}{\Delta_{\theta}} \d \theta^2 + \frac{\Delta_{\theta} \sin^2 \theta}{\Xi^2 \rho^2} \left[ a \d t - (r^2 + a^2) \d \phi \right]^2,
\label{Kerr-AdS-BL}
\eeq
where
\bearr
\rho^2          &=& r^2 + a^2 \cos^2 \theta, \\
\Delta_r        &=& (r^2 + a^2) \left( 1 + r^2 \right) - 2mr, \\
\Delta_{\theta} &=& 1 - a^2 \cos^2 \theta, \\
\Xi             &=& 1 - a^2.
\eearr
In order to make contact with the general form of the metric of the fluid/gravity map, it is interesting to write the metric in Eddington-Finkelstein-like coordinates. Consider the following transformation:
\beq
v = \frac{1}{\Xi} t + r^{\ast}, \qquad \varphi = \phi + \tilde{r},
\eeq
where $r^{\ast}$, $\tilde{r}$ are determined by
\beq
\frac{\d r^{\ast}}{\d r} = \frac{r^2 + a^2}{\Delta_r}, \qquad \frac{\d \tilde{r}}{\d r} = \frac{a \Xi}{\Delta_r}.
\eeq
Note that both equations can be integrated to give $r^{\ast}(r)$ and $\tilde{r}(r)$, since the RHS in both cases is a function of $r$ only. We therefore have
\beq
\d v = \frac{1}{\Xi} \d t + \frac{r^2 + a^2}{\Delta_r} \d r, \qquad \d \varphi = \d \phi + \frac{a \Xi}{\Delta_r} \d r.
\eeq
We now use these relations to change from $\{ t, r, \theta, \phi \}$ to $\{ v, r, \theta, \varphi \}$. The metric we obtain by doing this is
\beq
\begin{split}
ds^2 = - \frac{\Delta_r - \Delta_{\theta} a^2 \sin^2 \theta}{\rho^2} \d v^2 + 2 \d v \d r + \frac{2a\sin^2 \theta}{\Xi \rho^2} \left[ \Delta_r - \Delta_{\theta} (r^2 + a^2) \right] \d v \d \varphi \\ - \frac{2 a \sin^2 \theta}{\Xi} \d r \d \varphi + \frac{\rho^2}{\Delta_{\theta}} \d \theta^2 + \frac{\sin^2 \theta}{\Xi^2 \rho^2} \left[ \Delta_{\theta} (r^2 + a^2)^2 - \Delta_r a^2 \sin^2 \theta \right] \d \varphi^2.
\end{split}
\eeq


Following the AdS/CFT prescription \cite{BalasubramanianKraus1999} to determine the boundary stress tensor, one finds \cite{AwadJohnson2000, BhattacharyyaEtAl2008b}
\beq
\langle T_{ab} \rangle = p (3 u_a u_b + g_{ab}),
\label{Kerr-stress-tensor}
\eeq
where
\beq
p = \frac{m}{8\pi}, \qquad u^a = \left(\frac{\del}{\del v}\right)^a
\eeq
and the boundary metric $g_{ab}$ is\footnote{This is conformal to the metric of the Einstein static universe.}
\beq
ds_3^2 = - \d v^2 + \frac{2a\sin^2 \theta}{\Xi} \d v \d \varphi + \frac{1}{\Delta_{\theta}} \d \theta^2 + \frac{\sin^2 \theta}{\Xi} \d \varphi^2.
\label{Kerr-boundary-metric}
\eeq
Thus, the stress tensor describes exactly a \emph{perfect} conformal fluid at rest with constant pressure $p$, and hence constant temperature $T$.


This result seems surprising because one might have expected higher order corrections to the perfect fluid, constructed, for example, from the vorticity of the fluid, which will be shown below to be non-zero. We therefore want to use the formalism developed in Section \ref{section-conf-fluids-3-dim} and Appendix \ref{GHP-like-section} to understand this simple case in view of the fluid/gravity correspondence, which should be valid for large $m$. The fluid velocity $u^a$ will of course be chosen as the timelike basis vector. The other basis vectors can be chosen as
\beq
m^a = \frac{ia\sin \theta}{\sqrt{2 \Delta_{\theta}}} \left( \frac{\del}{\del v} \right)^a + \sqrt{\frac{\Delta_{\theta}}{2}} \left( \frac{\del}{\del \theta} \right)^a + \frac{i \Xi}{\sqrt{2 \Delta_{\theta}}\sin \theta} \left( \frac{\del}{\del \varphi} \right)^a
\eeq
and its complex-conjugate $\mbar^a$. In this basis, only the following connection components (see Appendix \ref{GHP-like-section}) are non-zero:
\beq
\omega = \tau = a \cos \theta, \qquad \kappa = \frac{\cos \theta}{\sqrt{2 \Delta_{\theta}}\sin \theta} \left( 1 - a^2 \cos 2 \theta \right).
\eeq
On the other hand, the only non-zero, independent curvature components are
\beq
\phi_0 = \Xi - 2 a^2 \cos 2 \theta, \qquad \phi_1 = \frac{ia \sqrt{\Delta_{\theta}}\sin \theta}{\sqrt{2}}.
\eeq
However, in this case all the $f$-components (see Appendix \ref{GHP-like-section}) vanish, thus allowing us to eliminate other quantities. In fact, one can verify explicitly that the following relations hold \emph{exactly}, which hold for any perfect fluid with vanishing shear---compare with Eqs. (\ref{scalar-elim-1}), (\ref{scalar-elim-2}):
\beq
\Dscr \omega = 0, \qquad \phi_1 = -i \delta \omega.
\label{Kerr-scalar-elim}
\eeq
Furthermore, the Bianchi identities reduce simply to
\beq
\Dscr \phi_0 = 0.
\eeq
Hence, for Kerr-AdS, we have only two independent scalars to consider, $\omega$ and $\phi_0$. These obey $\Dscr \omega = \Dscr \phi_0 = 0$, and the equations of motion are simply
\beq
\Dscr p = 0, \qquad \delta p = 0.
\eeq
Acting on a scalar of weight $(w,s)$, the commutators then reduce to
\bearr
(\Dscr \delta - \delta \Dscr)Q &=& i \omega \delta Q - i s Q \delta \omega, \\
(\Dscr \deltabar - \deltabar \Dscr)Q &=& - i \omega \deltabar Q - i s Q \deltabar \omega, \\
(\delta \deltabar - \deltabar \delta)Q &=& -2i \omega \Dscr Q - s Q \phi_0.
\eearr

We can now proceed as in the RT case: we classify the scalars built from $\omega$ and $\phi_0$ and their derivatives that could play a role in derivative expansions. In particular, we want to understand why the stress tensor describes \emph{exactly} a perfect fluid in this case. This can only happen if all non-zero contributions to the dissipative components $\pi_2,\bar{\pi}_2$ cancel at all orders. This conceivably allows us to determine additional transport coefficients. 

We begin by noting that we can use the commutators to bring any $\Dscr$-derivative to act directly on $\omega$ or $\phi_0$, at the expense of picking up terms that are products of scalars with fewer derivatives. Since $\Dscr \omega = \Dscr \phi_0 = 0$, we can ignore scalars containing $\Dscr$-derivatives and consider as fundamental objects only $\omega$, $\phi_0$, $\delta$, $\deltabar$.

There is a single one-derivative scalar in this case, $\omega$ (spin 0). At two derivatives, the scalars in (\ref{2-deriv-scalars}) reduce to
\beq
\begin{array}{lll}
\mbox{spin 0:} & \phi_0, & \omega^2 \\
\mbox{spin 1:} & \delta \omega. & {}
\end{array}
\eeq
We find the first possible contribution to $\pi_2$ at three derivatives only---cf. Eq. (\ref{3-deriv-scalars}):
\beq
\begin{array}{llll}
\mbox{spin 0:} & \delta \deltabar \omega, & \omega \phi_0, & \omega^3 \\
\mbox{spin 1:} & \delta \phi_0, & \omega \delta \omega & {} \\
\mbox{spin 2:} & \delta^2 \omega. & {} & {}
\end{array}
\label{3-deriv-scalars-Kerr}
\eeq
However, it turns out that, in this case, $\delta^2 \omega = 0$. One can show this is a consequence of conformal flatness of the metric (\ref{Kerr-boundary-metric}).\footnote{
This could also be understood as follows. The bulk solution is invariant under $t \rightarrow -t$, $\phi \rightarrow -\phi$ which acts on the basis vectors as a PT transformation. So $T_{ab}$ must be invariant under PT for this flow. Hence any PT violating term in $T_{ab}$ must either vanish for this flow or its coefficient must vanish (in which case the term is absent for all flows). The former happens for the term in $\pi_2$ proportional to $\delta^2 \omega$.}
 For any perfect fluid following a shear-free flow, such that Eqs. (\ref{Kerr-scalar-elim}) hold, vanishing of the Cotton tensor of the background metric is equivalent to
\bearr
\delta \deltabar \omega + \omega \phi_0 + 2 \omega^3 &=& 0, \label{conf-flat-cond-1} \\
\delta \phi_0 + 8 \omega \delta \omega &=& 0, \label{conf-flat-cond-2} \\
\delta^2 \omega &=& 0. \label{conf-flat-cond-3}
\eearr
One can verify that these three equations are satisfied for the Kerr-AdS case. Hence, we need to go at least to fourth order in derivatives to find a non-zero contribution to the stress tensor.

It is not difficult to classify the four-derivative scalars built solely from $\omega,\phi_0,\delta,\deltabar$. We obtain
\beq
\begin{array}{lllllll}
\mbox{spin 0:} & \omega^4, & \omega^2 \phi_0, & \omega \delta \deltabar \omega, & \delta \omega \deltabar \omega, & \phi_0^2, & \delta \deltabar \phi_0 \\
\mbox{spin 1:} & \omega^2 \delta \omega, & \delta^2 \deltabar \omega, & \phi_0 \delta \omega, & \omega \delta \phi_0 & {} & {} \\
\mbox{spin 2:} & \omega \delta^2 \omega, & (\delta \omega)^2, & \delta^2 \phi_0 & {} & {} & {}
\end{array}
\eeq
We know that $\delta^2 \omega = 0$, but
\beq
\delta \omega = - \frac{a \sqrt{\Delta_{\theta}} \sin \theta}{\sqrt{2}}.
\eeq
Hence $(\delta \omega)^2 \neq 0$, and this contribution must then be cancelled by the term $\delta^2 \phi_0$ in the stress tensor. In fact, using Eqs. (\ref{conf-flat-cond-2}), (\ref{conf-flat-cond-3}), we find
\beq
\delta^2 \phi_0 + 8 (\delta \omega)^2 = 0.
\eeq
The non-zero contribution to the stress tensor at fourth order in derivatives would then be
\bearr
\pi_2^{(4)} &=& p^{-1/3} \left[ C_{(\delta \omega)^2} (\delta \omega)^2 + C_{\delta^2 \phi_0} \delta^2 \phi_0 \right] \nonumber \\
            &=& p^{-1/3} \left[ C_{(\delta \omega)^2} - 8 C_{\delta^2 \phi_0} \right] (\delta \omega)^2.
\eearr
Since this must vanish, the relation
\beq
C_{(\delta \omega)^2} = 8 C_{\delta^2 \phi_0}
\eeq
must hold. But we have determined $C_{\delta^2 \phi_0}$ using the RT solution, Eq. (\ref{4-deriv-transport-coeff-RT}), and hence this allows us to determine
\beq
C_{(\delta \omega)^2} = \frac{1}{3\pi (16\pi)^{1/3}}.
\eeq

%
%

\section{Further comments and outlook}
\label{conclusions}

We investigated above the CFT interpretation of RT spacetimes. With a suitable choice of timelike congruence in the conformal boundary (i.e. a choice of frame), we found that the expectation value of the CFT energy-momentum tensor can be put exactly in the form
\beq
\label{CFTalgspec}
\langle T_{ab} \rangle = p_0 \left( 3 v_a v_b + g_{ab} \right) + \frac{1}{8\pi} Z_{(ab)},
\eeq
where $Z_{ab}$ is a three-derivative object built from the curvature, more specifically the Cotton tensor, of the boundary geometry, Eq. (\ref{Z-tensor}). This expression is local in the boundary metric, a property that can be traced back to the fact that the bulk spacetime is algebraically special. The leading order part has the form of a perfect conformal fluid at rest with constant pressure $p_0$ (equivalently constant temperature) flowing without shear or rotation in the background given by the conformal boundary of the RT spacetime, Eq. (\ref{boundary}). The three-derivative curvature term ensures that this energy-momentum tensor is conserved, as long as the RT equation is satisfied, i.e. as long as the bulk is really a member of the RT class.

A larger class of algebraically special solutions in $3+1$ dimensions is obtained by dropping the rotation-free condition defining RT solutions, i.e., we consider solutions with a shear-free null geodesic congruence for which the expansion and rotation are both non-vanishing. In this case, the vacuum Einstein equations can still be integrated, in the sense that all the dependence of the metric on a ``radial'' coordinate (affine parameter along the null congruence) is known, whereas the dependence on the ``boundary coordinates'' occurs through a few functions satisfying certain PDEs.

Such a general algebraically special metric can be put in the form \cite{ExactSolnsBook2003, KaigorodovTimofeev1996}
\beq
ds^2 = -2 (\d u + L \d \zeta + \Lbar \d \zetabar) \left[ \d r + W \d \zeta + \Wbar \d \zetabar + H (\d u + L \d \zeta + \Lbar \d \zetabar) \right] + \frac{2(r^2 + \Sigma^2)}{P^2} \d \zeta \d \zetabar.
\eeq
Here, $L = L(u,\zeta,\zetabar)$ is a complex function, $P = P(u,\zeta,\zetabar)$ is real, and
\beq
2i\Sigma = P^2 (\delbar L - \del \Lbar),
\eeq
\beq
W = i \del \Sigma - (r + i\Sigma) \del_u L,
\eeq
\beq
H = P^2 \mathrm{Re} \left[ \del (\delbar \ln P - \del_u \Lbar \right] - r \del_u \ln P - \frac{mr + M \Sigma}{r^2 + \Sigma^2} + \frac{r^2}{2} + \frac{5\Sigma^2}{2},
\eeq
with
\beq
\del = \del_{\zeta} - L\del_u, \quad \delbar = \del_{\zetabar} - \Lbar \del_u,
\eeq
and $m = m(u,\zeta,\zetabar)$, $M = M(u,\zeta,\zetabar)$ are real functions. Note that $\Sigma$ is also real. Furthermore, the RT solution is recovered when $L = 0$, in which case $\Sigma = 0$, $W = 0$ and $2H = \Phi$.

The vacuum Einstein equations have been integrated to give the complete dependence of the metric on the radial coordinate $r$. The dependence on the other coordinates is determined by a set of nonlinear PDEs given in Ref. \cite{KaigorodovTimofeev1996}.

Following the same procedure as for the RT case, we can consider the time-reversed solution with $u = -t$ and choose the conformal factor such that the boundary metric is
\beq
ds_3^2 = - \left[ \d t -L(t,\zeta,\zetabar) \d \zeta - \Lbar (t,\zeta,\zetabar) \d \zetabar \right]^2 + \frac{2}{P (t,\zeta,\zetabar)^2} \d \zeta \d \zetabar.
\label{twisting-boundary-metric}
\eeq
By taking again the velocity to be
\beq
v^a = \left( \frac{\del}{\del t} \right)^a,
\eeq
together with
\beq
\rho_0 = 2p_0 = \frac{m(t,\zeta,\zetabar)}{4\pi},
\eeq
we find that the AdS/CFT prescription for the boundary energy-momentum tensor yields precisely the same result as in the RT case, Eq. (\ref{CFTalgspec}), where $Z_{ab}$ is the same three-derivative curvature term as we had before, Eq. (\ref{Z-tensor}). Thus, with this choice of frame, the CFT state again assumes the form of a conformal fluid such that the only correction to the perfect fluid is a three-derivative object constructed from the curvature of the background geometry (\ref{twisting-boundary-metric}). In this case, the perfect fluid part has a velocity which is shear-free but has now a non-zero rotation. Notice that the Kerr-AdS solution is a special member of this class. The conformal boundary of Kerr-AdS is conformally flat, $C_{abc} = 0$, so in this case we recover the perfect fluid result of Section \ref{Kerr-AdS-section}, Eq. (\ref{Kerr-stress-tensor}).

We can summarize the above results in the following statement. \emph{In the conformal boundary of the general expanding, algebraically special spacetime in $3+1$ dimensions, there exists a shear-free timelike congruence. Choosing this as a reference frame, the dual CFT state corresponding to this general bulk spacetime is described by a conformal fluid whose energy-momentum tensor takes the form} (\ref{CFTalgspec}). This is a local function of the boundary metric $g_{ab}$, which is a consequence of the bulk being algebraically special.

We also studied above the fluid/gravity interpretation of RT and Kerr-AdS solutions, which enabled us to constrain a few transport coefficients at order higher than 2. In order to do this, we introduced in Section \ref{section-conf-fluids-3-dim} and Appendix \ref{GHP-like-section} a new formalism for studying conformal fluids in $2+1$ dimensions. The latter simplifies the manipulations and classifications for it involves dealing with scalar fields and partial derivatives only. As an illustration of its usefulness, we were easily able to classify in Section \ref{section-conf-fluids-3-dim} all independent, three-derivative objects that could contribute to the stress tensor at third order, Eq. (\ref{3-deriv-scalars}). In fact, it is not much more difficult to go beyond third order and obtain the complete classification also at the four-derivative level, although such results are not particularly useful for the contents of this paper.

Nevertheless, accomplishing a higher-order classification might be useful for other purposes. For instance, one might want to follow the procedure described in Ref. \cite{Bhattacharyya2012} to constrain some higher-order transport coefficients. By writing down the most general entropy current up to three derivatives and imposing its divergence (at fourth order in derivatives) to be non-negative, Ref. \cite{Bhattacharyya2012} obtained some constraints amongst transport coefficients of a general fluid flow at second order. However, as discussed there, all such constraints are automatically satisfied in the conformal case. Using our formalism, we can follow the same procedure (the relevant scalars were determined in Section \ref{section-conf-fluids-3-dim}) and recover these results in a more straightforward calculation. Of course, in our method we deal only with the three-dimensional conformal case, and hence no constraints on the energy-momentum tensor are obtained, only the form of the entropy current itself is constrained. It is then clear, as discussed in \cite{Bhattacharyya2012}, that if one wants to constrain the stress tensor for a conformal fluid one must go to higher orders in derivatives. It seems to us that the formalism introduced here can provide a great deal of simplification in this task.

As the GHP formalism has recently been generalized to higher dimensions \cite{DurkeeEtAl2010}, one might wish to extend the formalism presented here to $d > 3$. It is conceivable that a suitable generalization of this method may prove itself useful in treating conformal fluid dynamics in general.

Finally it is natural to ask whether there are higher-dimensional analogues of the algebraically special spacetimes studied here. Higher-dimensional Robinson-Trautman spacetimes, defined by the existence of an expanding shear-free, rotation-free null geodesic congruence, were studied in Ref. \cite{Podolsky:2006du}. For non-vanishing mass parameter (our $m$) it was found that the only such spacetime is the Schwarzschild-AdS solution. So higher-dimensional RT solutions do not exhibit the interesting time-dependence present in $3+1$ dimensions. In $4+1$ dimensions, a full classification of algebraically special solutions for which the preferred null direction is expanding and rotation-free was given in Ref. \cite{Reall:2012ih}. Most solutions are Kaluza-Klein or warped product spacetimes involving $3+1$ dimensional RT solutions. But a few other solutions were discovered and it would be interesting to investigate their AdS/CFT interpretation.

%
%

\section*{Acknowledgments}

We are grateful to Don Marolf and Mukund Rangamani for useful discussions. G.B.F. is supported by CAPES, grant no. 0252/11-5. HSR is supported by the European Research Council grant no. ERC-2011-StG 279363-HiDGR.

%
%
\fancyhead[RE,LO]{}
\appendix

%
%

\section{Weyl-covariant formalism}
\label{Weyl-cov-formalism}

The Weyl-covariant formalism introduced in Ref. \cite{Loganayagam2008} is particularly appropriate to study conformal fluids, as it automatically incorporates Weyl covariance. Here we summarize the notation and conventions used in this paper, referring the reader to \cite{Loganayagam2008} for more details.

Consider a tensor field $\mathcal{Q}^{a \ldots}{}_{b \ldots}$. We say that it is \emph{conformally covariant of weight} $w$ if, under a conformal transformation
\beq
g_{ab} \rightarrow \tilde{g}_{ab} = \Omega^2 g_{ab},
\eeq
it transforms homogeneously in the form
\beq
\tilde{\mathcal{Q}}^{a \ldots}{}_{b \ldots} = \Omega^w \mathcal{Q}^{a \ldots}{}_{b \ldots}.
\eeq
Thus the metric has weight $w = 2$. By introducing a Weyl connection $\calA_a$ transforming as
\beq
\tilde{\calA}_a = \calA_a + \nabla_a \ln \Omega,
\eeq
one can define a Weyl covariant derivative $\calD_a$ by
\bearr
\calD_c \mathcal{Q}^{a \ldots}{}_{b \ldots} &=& \nabla_c \mathcal{Q}^{a \ldots}{}_{b \ldots} - w \calA_c \mathcal{Q}^{a \ldots}{}_{b \ldots} \nonumber \\
                                            &\quad& + (g_{cd} \calA^a - \delta^a_c \calA_d - \delta^a_d \calA_c) \mathcal{Q}^{d \ldots}{}_{b \ldots} + \ldots \nonumber \\
                                            &\quad& - (g_{cb} \calA^d - \delta^d_c \calA_b - \delta^d_b \calA_c) \mathcal{Q}^{a \ldots}{}_{b \ldots} - \ldots .
\eearr
One can then verify that $\calD_c \mathcal{Q}^{a \ldots}{}_{b \ldots}$ is also conformally covariant with weight $w$. Furthermore, $\calD_a$ is metric-compatible, $\calD_a g_{bc} = 0$.

One can then define a Riemann curvature tensor by considering the commutator of two Weyl covariant derivatives. For example, if $X^a$ is a conformally covariant vector field of weight $w$, one finds
\beq
(\calD_a \calD_b - \calD_b \calD_a) X^c = \calR^c{}_{dab} X^d - w \calF_{ab} X^c,
\label{Weyl-Ricci-identity}
\eeq
where
\beq
\calF_{ab} = \nabla_a \calA_b - \nabla_b \calA_a = \tilde{\calF}_{ab}
\eeq
and
\beq
\calR_{cdab} = R_{cdab} - g_{cd} \calF_{ab} - 4 \delta^e_{[c} g_{d][a} \delta^f_{b]} \left( \nabla_f \calA_e + \calA_f \calA_e - \frac{\calA^2}{2} g_{fe} \right) = \Omega^{-2} \tilde{\calR}_{cdab}.
\label{curly-Riemann}
\eeq
It is then straightforward to define a corresponding Weyl-covariant Ricci tensor,
\beq
\calR_{ab} = \calR^c{}_{acb} = R_{ab} - \calF_{ab} + (d-2) \left( \nabla_b \calA_a + \calA_b \calA_a - \calA^2 g_{ba} \right) + g_{ab} \nabla_c \calA^c = \tilde{\calR}_{ab},
\label{curly-Ricci}
\eeq
and a Weyl-covariant Ricci scalar,
\beq
\calR = R + 2(d-1) \nabla_a \calA^a - (d-1)(d-2) \calA^2 = \Omega^{2} \tilde{\calR}.
\eeq

It is important to notice that the Weyl-covariant curvature tensors do not possess the same set of symmetries as the conventional curvature tensors constructed from the metric. Some useful symmetry relations are
\bearr
\calR_{(ab)cd} &=& -g_{ab} \calF_{cd}, \\
\calR_{abcd} - \calR_{cdab} &=& 4 \delta^e_{[a} g_{b][c} \delta^f_{d]} \calF_{ef} - g_{ab} \calF_{cd} + g_{cd} \calF_{ab}, \\
\calR_{ab} - \calR_{ba} &=& - d \calF_{ab}. \label{curly-Ricci-symmetry}
\eearr
The relation $\calR^a{}_{[bcd]} = 0$ holds, however. Furthermore, the curvature tensors above also obey various Bianchi identities and their contractions:
\bearr
\calD_{[a} \calF_{bc]} &=& 0, \\
\calD_{[a} \calR^b{}_{|c|de]} &=& 0, \\
\calD_a \calR^a{}_{bcd} - \calD_c \calR_{bd} + \calD_d \calR_{bc} &=& 0, \\
\calD_a \left( \calR^{ab} - \frac{1}{2} \calR g^{ab} + \calF^{ab} \right) &=& 0.
\eearr

When one is considering a conformal fluid on the background with metric $g_{ab}$, there is a natural, preferred vector field, namely the fluid velocity $u^a$. This can be used to fix the ambiguity in $\calA_a$. In particular, when working in Landau frame, it is natural to impose
\beq
u^a \calD_a u_b = 0, \qquad g^{ab} \calD_a u_b = 0,
\eeq
so that
\beq
\calD_a u_b = \sigma_{ab} + \omega_{ab}
\eeq
is transverse and traceless. It turns out that these conditions uniquely determine $\calA_a$ to be
\beq
\calA_a = a_a - \frac{\theta}{d-1} u_a,
\label{Weyl-field-strength}
\eeq
where
\beq
a^a = u^b \nabla_b u^a
\eeq
is the \emph{acceleration}, and
\beq
\theta = \nabla_a u^a
\eeq
is the \emph{expansion}.

In three dimensions, the Cotton tensor (\ref{Cotton-tensor}) plays an important role. It is conformally invariant and its vanishing is equivalent to conformal flatness. This can be written in terms of the Weyl-covariant formalism as
\beq
C_{abc} = \calD_c \left( \calR_{ba} - \frac{1}{4} \calR g_{ba} \right) - \calD_b \left( \calR_{ca} - \frac{1}{4} \calR g_{ca} \right) + 2 \calD_a \calF_{bc}.
\eeq

%
%

\section{A GHP formalism for fluids in 2+1 dimensions}
\label{GHP-like-section}

Motivated by the GHP formalism \cite{GerochHeldPenrose1973} that is used to study algebraically special spacetimes, and the Weyl-covariant formalism defined in the previous Appendix, we can develop a new formalism that is particularly useful to study conformal fluids in $2+1$ dimensions. In GHP, one has two preferred null directions that one chooses as null basis vectors. In contrast, in fluid dynamics there is a preferred \emph{timelike} congruence instead given by the fluid velocity. This timelike vector field can be chosen as one of the basis vectors at every point. One can then complete the basis with spacelike vector fields to form an orthonormal basis. Every tensor field can then be projected along the basis to form scalar fields. As the spatial directions can be rotated at will, one is only interested in those scalars transforming homogeneously under such rotations. Moreover, if one is interested in conformal fluids, then it is natural to restrict oneself to scalars that furthermore transform homogeneously under conformal rescalings as well. In general, derivatives of such scalars will not possess the same transformation properties, even when projected along the basis. However, one can correct derivatives with connection terms to deal automatically with objects having the desired transformation properties.

Let us develop the ideas above in detail for the case of $2+1$ dimensions. In this case, take a basis $\{ u^a, m_{(1)}^a, m_{(2)}^a \}$, where $u^a$ is unit timelike and the $m_{(i)}^a$ are unit spacelike, orthogonal to each other and to $u^a$:
\beq
g_{ab} u^a u^b = -1, \qquad g_{ab} m_{(i)}^a m_{(j)}^b = \delta_{ij},
\eeq
with all other inner products zero. The metric is then
\beq
g_{ab} = -u_a u_b + \left( m_{(1)} \right)_a \left( m_{(1)} \right)_b + \left( m_{(2)} \right)_a \left( m_{(2)} \right)_b.
\eeq
We will sometimes find it useful to assume that this orthonormal basis is also right-handed,
\beq
\epsilon_{abc} u^a m_{(1)}^b m_{(2)}^c = 1.
\eeq
Now define the complex vectors
\beq
m^a = \frac{1}{\sqrt{2}} \left( m_{(1)}^a + i m_{(2)}^a \right), \qquad \mbar^a = \frac{1}{\sqrt{2}} \left( m_{(1)}^a - i m_{(2)}^a \right).
\eeq
We then have
\beq
g_{ab} m^a m^b = g_{ab} \mbar^a \mbar^b = 0, \qquad g_{ab} m^a \mbar^b = 1,
\eeq
and the metric is simply
\beq
g_{ab} = -u_a u_b + m_a \mbar_b + \mbar_a m_b.
\eeq

We are interested in two fundamental transformations of the basis vectors. A \emph{conformal transformation}
\beq
g_{ab} \rightarrow \tilde{g}_{ab} = \Omega^2 g_{ab}
\eeq
rescales the basis according to
\beq
u^a \rightarrow \tilde{u}^a = \frac{1}{\Omega} u^a, \qquad m^a \rightarrow \tilde{m}^a = \frac{1}{\Omega} m^a.
\eeq
A \emph{rotation} of the spatial directions is given by
\beq
u^a \rightarrow \tilde{u}^a = u^a, \qquad m^a \rightarrow \tilde{m}^a = e^{i\lambda} m^a.
\eeq
The general transformation is therefore
\beq
\tilde{u}^a = \frac{1}{\Omega} u^a, \qquad \tilde{m}^a = \frac{e^{i\lambda}}{\Omega} m^a,
\label{conf-spin}
\eeq
and we want to see the corresponding transformation induced on other objects, in particular scalars.

It is more convenient, however, to treat the general case and specialize to scalars later. The connection components are encoded in
\beq
L_{ab} \equiv \nabla_a u_b, \qquad M_{ab} \equiv \nabla_a m_b, \qquad \bar{M}_{ab} \equiv \nabla_a \mbar_b.
\eeq
Not all components are independent, however, due to the normalization and orthogonality conditions. We have
\beq
L_{ab} u^b = 0, \qquad M_{ab} m^b = 0, \qquad \bar{M}_{ab} \mbar^b = 0,
\eeq
and
\beq
L_{ab} m^b + M_{ab} u^b = 0, \qquad L_{ab} \mbar^b + \bar{M}_{ab} u^b = 0, \qquad M_{ab} \mbar^b + \bar{M}_{ab} m^b = 0.
\eeq
Thus, we can find nine independent components in total, which are summarized in Table \ref{conn-compts} below. In general, the connection components transform inhomogeneously under the general transformation (\ref{conf-spin}). In fact, one finds
\bearr
\tilde{L}_{ab} &=& \Omega \left( L_{ab} - u_a \nabla_b \ln \Omega + g_{ab} u^c \nabla_c \ln \Omega \right), \\
\tilde{M}_{ab} &=& \Omega e^{i\lambda} \left( M_{ab} + i m_b \nabla_a \lambda - m_a \nabla_b \ln \Omega + g_{ab} m^c \nabla_c \ln \Omega \right).
\eearr
The only components that do transform homogeneously are the two components of the shear, $\sigma, \bar{\sigma}$, and the rotation $\omega$. These transformation properties are also summarized in Table \ref{conn-compts}.

\begin{table}[ht]

\caption{Connection components}

\begin{center}

\begin{tabular}{l|l|l}

Coefficient & Transformation & Interpretation \\
\hline

$a = L_{ab} u^a m^b$, $\bar{a} = L_{ab} u^a \mbar^b$ & $\tilde{a} = \frac{e^{i\lambda}}{\Omega} \left( a + m^a \nabla_a \ln \Omega \right)$ & Acceleration of $u^a$ \\

$\sigma = L_{ab} m^a m^b$, $\bar{\sigma} = L_{ab} \mbar^a \mbar^b$ & $\tilde{\sigma} = \frac{e^{2i\lambda}}{\Omega} \sigma$ & Shear of $u^a$ \\

$\theta = L_{ab} (m^a \mbar^b + \mbar^a m^b) = \bar{\theta}$ & $\tilde{\theta} = \frac{1}{\Omega} \left( \theta + 2 u^a \nabla_a \ln \Omega \right)$ & Expansion of $u^a$ \\

$\omega = \frac{i}{2} L_{ab} (m^a \mbar^b - \mbar^a m^b) = \bar{\omega}$ & $\tilde{\omega} = \frac{1}{\Omega} \omega$ & Rotation or vorticity of $u^a$ \\

$\tau = i M_{ab} u^a \mbar^b = \bar{\tau}$ & $\tilde{\tau} = \frac{1}{\Omega} \left( \tau - u^a \nabla_a \lambda \right)$ & Transport of $m^a, \mbar^a$ along $u^a$ \\

$\kappa = M_{ab} m^a \mbar^b$, $\bar{\kappa} = - M_{ab} \mbar^a \mbar^b$ & $\tilde{\kappa} = \frac{e^{i\lambda}}{\Omega} \left( \kappa + i m^a \nabla_a \lambda + m^a \nabla_a \ln \Omega \right)$ & Non-geodesity of $m^a,\mbar^a$

\end{tabular}

\end{center}

\label{conn-compts}

\end{table}

Although there are some connection components that do transform homogeneously under (\ref{conf-spin}), this will not be the case when we take derivatives, in general. Following the same ideas as in GHP and the Weyl-covariant formalism of the previous Appendix, we would like to define a new derivative operator that preserves transformation properties under (\ref{conf-spin}). We will say that a tensor field $Q^{a \ldots}{}_{b \ldots}$ has \emph{conformal weight} $w$ and \emph{spin weight} $s$, abbreviated weight $(w,s)$, if, under the transformation (\ref{conf-spin}), it transforms as
\beq
\tilde{Q}^{a \ldots}{}_{b \ldots} = \Omega^w e^{is\lambda} Q^{a \ldots}{}_{b \ldots}.
\eeq
Of course, most tensors of relevance, e.g. the metric, energy-momentum tensor, curvature tensors, will have spin weight $s = 0$, as they are independent of the choice of $m^a,\mbar^a$. However, one may still encounter tensor fields that can be rotated under (\ref{conf-spin}), in particular $m^a,\mbar^a$ themselves and their outer products.

Suppose we can find two one-forms $A_a, B_a$ that transform as
\beq
\tilde{A}_a = A_a + \nabla_a \ln \Omega, \qquad \tilde{B}_a = B_a + \nabla_a \lambda
\eeq
under (\ref{conf-spin}). Then we can define a new derivative operator acting on a tensor field of weight $(w,s)$ by
\bearr
D_c Q^{a \ldots}{}_{b \ldots} &=& \nabla_c Q^{a \ldots}{}_{b \ldots} - (w A_c + i s B_c) Q^{a \ldots}{}_{b \ldots} \nonumber \\
                                            &\quad& + (g_{cd} A^a - \delta^a_c A_d - \delta^a_d A_c) Q^{d \ldots}{}_{b \ldots} + \ldots \nonumber \\
                                            &\quad& - (g_{cb} A^d - \delta^d_c A_b - \delta^d_b A_c) \mathcal{Q}^{a \ldots}{}_{b \ldots} - \ldots,
\eearr
which also has weigth $(w,s)$:
\beq
\tilde{D}_c \tilde{Q}^{a \ldots}{}_{b \ldots} = \Omega^w e^{is\lambda} D_c Q^{a \ldots}{}_{b \ldots}.
\eeq
Note that $D_c Q^{a \ldots}{}_{b \ldots} = \calD_c Q^{a \ldots}{}_{b \ldots} - isB_c Q^{a \ldots}{}_{b \ldots}$, where $\calD_a$ is the Weyl-covariant derivative defined in the previous Appendix. Using only the connection components of Table \ref{conn-compts}, we can determine $A_a$ and $B_a$ to be
\bearr
A_a &=& - \frac{\theta}{2} u_a + \bar{a} m_a + a \mbar_a, \\
B_a &=& \tau u_a - i (\bar{a} - \bar{\kappa}) m_a + i (a - \kappa) \mbar_a.
\eearr
Note that $A_a = - \frac{\theta}{2} u_a + a_a$, in agreement with the conventional choice determined in the previous Appendix, Eq. (\ref{Weyl-field-strength}). In particular, we have
\bearr
D_a u_b     &=& (i\omega m_a + \sigma \mbar_a) \mbar_b + (\bar{\sigma} m_a - i\omega \mbar_a) m_b, \\
D_a m_b     &=& (i\omega m_a + \sigma \mbar_a) u_b, \\
D_a \mbar_b &=& (\bar{\sigma} m_a - i\omega \mbar_a) u_b.
\eearr

We can now define curvature tensors as in the Weyl-covariant approach, by considering the commutator of two derivatives. For example, if $X^a$ is now a vector field of weight $(w,s)$, one finds
\beq
(D_c D_d - D_d D_c) X^a = \calR^a{}_{bcd} X^b - (w F_{cd} + is H_{cd}) X^a,
\eeq
where $\calR^a{}_{bcd}$ is the Weyl-covariant Riemann tensor of the previous Appendix, Eq. (\ref{curly-Riemann}), and
\beq
F_{ab} = \nabla_a A_b - \nabla_b A_a, \qquad H_{ab} = \nabla_a B_b - \nabla_b B_a,
\eeq
both of which are invariant under (\ref{conf-spin}).

Now we want to take components of the various tensor fields by projecting along the basis vectors thus dealing with scalars only. Hence, if $Q$ is a scalar of weight $(w,s)$, we define the derivatives
\beq
\Dscr Q \equiv u^a D_a Q, \qquad \delta Q \equiv m^a D_a Q, \qquad \deltabar Q \equiv \mbar^a D_a Q.
\eeq
Explicitly, we have
\bearr
\Dscr Q &=& u^a \nabla_a Q - \left( \frac{w}{2} \theta - i s \tau \right) Q, \label{Dscr-def} \\
\delta Q &=& m^a \nabla_a Q - \left[ w a + s (\kappa - a) \right] Q, \label{delta-def} \\
\deltabar Q &=& \mbar^a \nabla_a Q - \left[ w \bar{a} + s (\bar{a} - \bar{\kappa}) \right] Q. \label{deltabar-def}
\eearr
We see that $\Dscr Q$, $\delta Q$ and $\deltabar Q$ have weights $(w-1,s)$, $(w-1,s+1)$ and $(w-1,s-1)$, respectively.

As we saw above, the only connection components that have definite conformal and spin weights are the shear and rotation, $\sigma, \bar{\sigma}, \omega$. On the other hand, all components of the curvature tensors corresponding to the derivative operator $D_a$ are of course scalars with definite weight. In three dimensions, the Weyl-covariant Riemann tensor is completely determined by the Weyl-covariant Ricci tensor and the ``field strength'' $F_{ab}$. In fact, it is convenient to deal with the symmetric and antisymmetric parts of the Ricci tensor separately. We define $\Phi_{ab} = \calR_{(ab)}$, so that
\beq
\calR_{ab} = \Phi_{ab} - \frac{3}{2} F_{ab},
\eeq
cf. Eq. (\ref{curly-Ricci-symmetry}). Then, one can show
\bearr
\calR_{abcd} &=& -g_{ab} F_{cd} + g_{ac} \left( \Phi_{bd} - \frac{1}{2} F_{bd} - \frac{1}{4} \Phi g_{bd} \right) - g_{ad} \left( \Phi_{bc} - \frac{1}{2} F_{bc} - \frac{1}{4} \Phi g_{bc} \right) \nonumber \\
             &\quad& \quad - g_{bc} \left( \Phi_{ad} - \frac{1}{2} F_{ad} - \frac{1}{4} \Phi g_{ad} \right) + g_{bd} \left( \Phi_{ac} - \frac{1}{2} F_{ac} - \frac{1}{4} \Phi g_{ac} \right)
\eearr
in three dimensions, where $\Phi = g^{ab} \Phi_{ab} = \calR$. The relevant curvature scalars are then summarized in Table \ref{curv-scalars}.

\begin{table}[ht]

\caption{Curvature scalars}

\begin{center}

\begin{tabular}{c|c}

Coefficient & Weight $(w,s)$ \\
\hline

$\phi_2 = \Phi_{ab} m^a m^b$ & (-2,2) \\

$\phi_1 = \Phi_{ab} u^a m^b$ & (-2,1) \\

$\phi_0 = \Phi_{ab} m^a \mbar^b = \bar{\phi_0}$ & (-2,0) \\

$\phi'_0 = \Phi_{ab} u^a u^b = \bar{\phi'_0}$ & (-2,0) \\

$\phi_{-1} = \Phi_{ab} u^a \mbar^b = \bar{\phi_1}$ & (-2,-1) \\

$\phi_{-2} = \Phi_{ab} \mbar^a \mbar^b = \bar{\phi_2}$ & (-2,-2) \\

$f_1 = F_{ab} u^a m^b$ & (-2,1) \\

$f_0 = i F_{ab} m^a \mbar^b = \bar{f_0}$ & (-2,0) \\

$f_{-1} = F_{ab} u^a \mbar^b = \bar{f_1}$ & (-2,-1) \\

$h_1 = H_{ab} u^a m^b$ & (-2,1) \\

$h_0 = i H_{ab} m^a \mbar^b = \bar{h_0}$ & (-2,0) \\

$h_{-1} = H_{ab} u^a \mbar^b = \bar{h_1}$ & (-2,-1)

\end{tabular}

\end{center}

\label{curv-scalars}

\end{table}

Not all the curvature scalars and the relevant connection components are independent, for they are related by Bianchi identities and the analogs of the Newman-Penrose equations. The latter are obtained by considering the Ricci identity
\beq
(D_c D_d - D_d D_c) X_b = - \calR_{abcd} X^a - (w F_{cd} + is H_{cd}) X_b,
\eeq
putting $X_a = u_a,m_a,\mbar_a$ and taking components along the basis vectors. One obtains
\bearr
\Dscr \sigma &=& \phi_2, \\
\Dscr \omega &=& \frac{1}{2} f_0, \\
\phi'_0 &=& - 2 (\sigma \bar{\sigma} - \omega^2), \\
\deltabar \sigma - i \delta \omega &=& \phi_1 - \frac{1}{2} f_1, \\
h_1 &=& i \left(\phi_1 + \frac{1}{2} f_1  \right), \\
h_0 &=& \phi_0 + \frac{1}{2} \phi'_0 + \sigma \bar{\sigma} - \omega^2,
\eearr
together with the complex-conjugate relations, when appropriate. The first two of these are equations for the propagation of shear and rotation, respectively, whereas the third is equivalent to Raychaudhuri's equation (describing the propagation of the expansion) in three dimensions. We can then see that all the $f$- and $h$-scalars, together with $\phi_0',\phi_2,\bar{\phi}_2$, can be eliminated in terms of other curvature and connection scalars. We can then consider the commutator of two derivatives acting on a scalar $Q$ of weight $(w,s)$,
\beq
(D_a D_b - D_b D_a) Q = - (w F_{ab} + is H_{ab}) Q,
\eeq
and take the nontrivial components along the basis to find
\bearr
(\Dscr \delta - \delta \Dscr) Q &=& i \omega \delta Q - \sigma \deltabar Q - 2w (\phi_1 - \deltabar \sigma + i \delta \omega) Q + s (2 \phi_1 - \deltabar \sigma + i \delta \omega) Q, \\
(\Dscr \deltabar - \deltabar \Dscr) Q &=& - i \omega \deltabar Q - \bar{\sigma} \delta Q - 2w (\bar{\phi}_1 - \delta \bar{\sigma} - i \deltabar \omega) Q - s (2 \bar{\phi}_1 - \delta \bar{\sigma} - i \deltabar \omega) Q, \\
(\delta \deltabar - \deltabar \delta) Q &=& -2i \omega \Dscr Q + (2iw \Dscr \omega - s \phi_0 )Q.
\eearr

The relevant Bianchi identities are
\beq
D_{[e} \calR_{|ab|cd]} = 0, \qquad D_{[a} F_{bc]} = 0, \qquad D_{[a} H_{bc]} = 0.
\eeq
Taking components gives the following equations:
\bearr
0 &=& \Dscr \left( \phi_1 - \frac{1}{2} f_1 \right) - \frac{1}{2} \delta (\phi'_0 - if_0) - \deltabar \phi_2 + \sigma \left( \bar{\phi}_1 + \frac{1}{2} \bar{f}_1 \right) + i \omega \left( \phi_1 + \frac{1}{2} f_1 \right), \\
0 &=& \Dscr \left( \phi_0 + \frac{1}{2} \phi'_0 \right) - \delta \left( \bar{\phi}_1 + \frac{1}{2} \bar{f}_1 \right) - \deltabar \left( \phi_1 + \frac{1}{2} f_1 \right) + \sigma \bar{\phi}_2 + \bar{\sigma} \phi_2 - \omega f_0, \\
0 &=& \Dscr f_0 -i \delta \bar{f}_1 +i \deltabar f_1, \\
0 &=& \Dscr h_0 -i \delta \bar{h}_1 +i \deltabar h_1.
\eearr
Using the eliminations provided by the Newman-Penrose equations, one finds that the Bianchi identities reduce to the following two non-redundant equations:
\bearr
0 &=& \deltabar \phi_1 - \delta \bar{\phi}_1 + \delta^2 \bar{\sigma} - \deltabar^2 \sigma - i (\Dscr^2 \omega - \delta \deltabar \omega - \deltabar \delta \omega), \\
0 &=& \Dscr \phi_0 - 2 \delta \bar{\phi}_1 - 2 \deltabar \phi_1 + \delta^2 \bar{\sigma} + \deltabar^2 \sigma + 4 \omega \Dscr \omega.
\eearr

%
%

%
%


\begin{thebibliography}{99}

\bibitem{RobinsonTrautman1962}
I. Robinson and A. Trautman, \emph{Some spherical gravitational waves in general relativity}, Proc. R. Soc. Lond. A \textbf{265}, 463-473 (1962).

\bibitem{BhattacharyyaEtAl2008a}
Sayantani Bhattacharyya, Veronika E. Hubeny, Shiraz Minwalla and Mukund Rangamani, \emph{Nonlinear Fluid Dynamics from Gravity}, JHEP \textbf{0802}, 045 (2008). arXiv:0712.2456

\bibitem{HubenyMinwallaRangamani2011}
Veronika E. Hubeny, Shiraz Minwalla and Mukund Rangamani, \emph{The fluid/gravity correspondence}, in: \emph{Black Holes in Higher Dimensions}, edited by Gary T. Horowitz, Cambridge University Press (2012). arXiv:1107.5780

\bibitem{Mukhopadhyay:2013gja} 
  A.~Mukhopadhyay, A.~C.~Petkou, P.~M.~Petropoulos, V.~Pozzoli and K.~Siampos,
  \emph{Holographic perfect fluidity, Cotton energy-momentum duality and transport properties,}
  arXiv:1309.2310 [hep-th].

\bibitem{GerochHeldPenrose1973}
R.P. Geroch, A. Held and R. Penrose, \emph{A space-time calculus based on pairs of null directions}, J. Math. Phys. \textbf{14}, 874-881 (1973).

\bibitem{AwadJohnson2000}
Adel M. Awad and Clifford V. Johnson, \emph{Holographic stress tensors for Kerr-AdS black holes}, Phys. Rev. D \textbf{61}, 084025 (2000). arXiv:hep-th/9910040

\bibitem{BhattacharyyaEtAl2008b}
Sayantani Bhattacharyya, R. Loganayagam, Ipsita Mandal, Shiraz Minwalla and Ankit Sharma, \emph{Conformal Nonlinear Fluid Dynamics from Gravity in Arbitrary Dimensions}, JHEP \textbf{0812}, 116 (2008). arXiv:0809.4272

\bibitem{ExactSolnsBook2003}
H. Stephani, D. Kramer, M. MacCallum, C. Hoenselaers and E. Herlt, \emph{Exact Solutions of Einstein's Field Equations} (second edition), Cambridge University Press (2003).

\bibitem{GriffithsPodolsky2012}
J.B. Griffiths and J. Podolsk\'{y}, \emph{Exact Space-Times in Einstein's General Relativity}, Cambridge University Press (2012).

\bibitem{Chrusciel1991}
P.T. Chrusciel, \emph{Semiglobal existence and convergence of solutions of the Robinson-Trautman (two-dimensional Calabi) equation}, Commun. Math. Phys. \textbf{137}, 289 (1991).

\bibitem{ChruscielSingleton1992}
P.T. Chrusciel and D.B. Singleton, \emph{Non-smoothness of event horizons of Robinson-Trautman black holes}, Commun. Math. Phys. \textbf{147}, 137-162 (1992).

\bibitem{BicakPodolsky1997}
Jiri Bicak and Jiri Podolsky, \emph{Global structure of Robinson-Trautman radiative space-times with a cosmological constant}, Phys. Rev. D \textbf{55}, 1985-1993 (1997). arXiv:gr-qc/9901018

\bibitem{BalasubramanianKraus1999}
V. Balasubramanian and P. Kraus, {\em A Stress Tensor for Anti-de Sitter Gravity}, Commun. Math. Phys. \textbf{208}, 413-428 (1999). arXiv:hep-th/9902121

\bibitem{Wald1977}
R.M. Wald, \emph{The Back Reaction Effect in Particle Creation in Curved Spacetime}, Commun. Math. Phys. \textbf{54}, 1-19 (1977)

\bibitem{BaierEtAl2008}
R. Baier, P. Romatschke, D.T. Son, A.O. Starinets and M.A. Stephanov, \emph{Relativistic viscous hydrodynamics, conformal invariance, and holography}, JHEP \textbf{0804}, 100 (2008). arXiv:0712.2451

\bibitem{Loganayagam2008}
R. Loganayagam, {\em Entropy Current in Conformal Hydrodynamics}, JHEP \textbf{0805}, 087 (2008). arXiv:0801.3701

\bibitem{VanRaamsdonk2008}
Mark Van Raamsdonk, \emph{Black Hole Dynamics From Atmospheric Science}, JHEP \textbf{0805}, 106 (2008). arXiv:0802.3224

\bibitem{CarrascoEtAl2012}
Federico Carrasco, Luis Lehner, Robert C. Myers, Oscar Reula and Ajay Singh, \emph{Turbulent flows for relativistic conformal fluids in 2+1 dimensions}, Phys. Rev. D \textbf{86}, 126006 (2012). arXiv:1210.6702

\bibitem{SaremiSon2011}
Omid Saremi and Dam Thanh Son, \emph{Hall viscosity from gauge/gravity duality}, JHEP \textbf{1204}, 091 (2012). arXiv:1103.4851

\bibitem{JensenEtAl2011}
Kristan Jensen, Matthias Kaminski, Pavel Kovtun, Ren\'{e} Meyer, Adam Ritz and Amos Yarom, \emph{Parity-Violating Hydrodynamics in 2+1 Dimensions}, JHEP \textbf{1205}, 102 (2012). arXiv:1112.4498

\bibitem{ChenEtAl2012}
Jiunn-Wei Chen, Shou-Huang Dai, Nien-En Lee and Debaprasad Maity, \emph{Novel Parity Violating Transport Coefficients in 2+1 Dimensions from Holography}, JHEP \textbf{1209}, 096 (2012). arXiv:1206.0850

\bibitem{Romatschke2010}
Paul Romatschke, \emph{Relativistic Viscous Fluid Dynamics and Non-Equilibrium Entropy}, Class. Quant. Grav. \textbf{27}, 025006 (2010). arXiv:0906.4787

\bibitem{Bhattacharyya2012}
Sayantani Bhattacharyya, \emph{Constraints on the second order transport coefficients of an uncharged fluid}, JHEP \textbf{1207}, 104 (2012). arXiv:1201.4654

\bibitem{BhattacharyyaEtAl2008c}
Sayantani Bhattacharyya, Veronika E. Hubeny, R. Loganayagam, Gautam Mandal, Shiraz Minwalla, Takeshi Morita, Mukund Rangamani and Harvey S. Reall, \emph{Local Fluid Dynamical Entropy from Gravity}, JHEP \textbf{0806}, 055 (2008). arXiv:0803.2526

\bibitem{KaigorodovTimofeev1996}
V.R. Kaigorodov and V.N. Timofeev, \emph{Algebraically Special Solutions of the Einstein Equations $R_{ij} = 6 \Lambda g_{ij}$}, Gravitation \& Cosmology \textbf{2} (2), 107-108 (1996).

\bibitem{DurkeeEtAl2010}
Mark Durkee, Vojt Pravda, Alena Pravdova and Harvey S. Reall, \emph{Generalization of the Geroch-Held-Penrose formalism to higher dimensions}, Class. Quant. Grav. \textbf{27}, 215010 (2010). arXiv:1002.4826

\bibitem{Podolsky:2006du} 
  J.~Podolsky and M.~Ortaggio,
  \emph{Robinson-Trautman spacetimes in higher dimensions},
  Class.\ Quant.\ Grav.\  {\bf 23}, 5785 (2006)
  [gr-qc/0605136].

\bibitem{Reall:2012ih} 
  H.~S.~Reall, A.~A.~H.~Graham and C.~P.~Turner,
  \emph{On algebraically special vacuum spacetimes in five dimensions},
  Class.\ Quant.\ Grav.\  {\bf 30}, 055004 (2013)
  [arXiv:1211.5957 [gr-qc]].

\end{thebibliography}

\end{document}